%% file: main.tex
\documentclass[letterpaper,twocolumn,10pt]{article}
\usepackage[dvipsnames]{xcolor}
\usepackage[T1]{fontenc}
    \usepackage{textcomp}
\usepackage{fontawesome5}

\usepackage{usenix-2020-09}

\usepackage{amsmath}
\usepackage{adjustbox}
\usepackage[toc,page]{appendix}
\usepackage{balance}
\usepackage{braket}
\usepackage{here}
\usepackage{lipsum}
\usepackage{listings}
\usepackage{svg}
\usepackage{tikz}
\usepackage[normalem]{ulem}
\useunder{\uline}{\ul}{}
\usepackage{widetext}

\definecolor{codegreen}{rgb}{0,0.6,0}
\definecolor{codegray}{rgb}{0.5,0.5,0.5}
\definecolor{codepurple}{rgb}{0.58,0,0.82}
\definecolor{backcolour}{rgb}{0.95,0.95,0.92}

\definecolor{color1}{RGB}{202, 60, 110}
\definecolor{color2}{RGB}{110, 183, 219}

\lstdefinestyle{mystyle}{
    backgroundcolor=\color{backcolour},   
    commentstyle=\color{codegreen},
    keywordstyle=\color{magenta},
    numberstyle=\tiny\color{codegray},
    stringstyle=\color{codepurple},
    basicstyle=\ttfamily\footnotesize,
    breakatwhitespace=false,         
    breaklines=true,                 
    captionpos=b,       
    frame=single,
    keepspaces=true,                 
    numbers=left,                    
    showspaces=false,                
    showstringspaces=false,
    showtabs=false,                  
    tabsize=2
}

\lstdefinelanguage{Ini}
{
    basicstyle=\ttfamily\small,
    columns=fullflexible,
    morecomment=[s][\color{Orchid}\bfseries]{[}{]},
    morecomment=[l]{\#},
    morecomment=[l]{;},
    commentstyle=\color{gray}\ttfamily,
    morekeywords={},
    otherkeywords={},
    keywordstyle={\color{black}\bfseries}
}

\lstdefinelanguage{Ned}
{
    basicstyle=\ttfamily\small,
    columns=fullflexible,
    morekeywords=[1]{network, submodules, connections},
    keywordstyle=[1]{\color{color1}\bfseries},
    morekeywords=[2]{QNode, HoM, ClassicalChannel, QuantumChannel},
    keywordstyle=[2]{\color{color2}\bfseries},
}

\lstdefinelanguage{Txt}
{
    basicstyle=\ttfamily\small,
    columns=fullflexible,
    morecomment=[l]{//},
    morecomment=[l]{;},
    commentstyle=\color{gray}\ttfamily,
    morekeywords={},
    otherkeywords={:},
    keywordstyle={\color{black}\bfseries}
}

\RequirePackage{array}
\RequirePackage{longtable}
\RequirePackage{hyperref}
\RequirePackage{multirow}
\RequirePackage{wrapfig}

\newcolumntype{L}[1]{>{\raggedright\let\newline\\\arraybackslash\hspace{0pt}}p{#1}}
\newcolumntype{M}[1]{>{\raggedright\let\newline\\\arraybackslash\hspace{0pt}}m{#1}}
\newcolumntype{C}[1]{>{\centering\let\newline\\\arraybackslash\hspace{0pt}}p{#1}}
\newcolumntype{R}[1]{>{\raggedleft\let\newline\\\arraybackslash\hspace{0pt}}p{#1}}

\RequirePackage{ifthen}

\newboolean{ShowComments}
\setboolean{ShowComments}{false}  
\ifthenelse{\boolean{ShowComments}}%
	{
		\newcommand{\ColorComment}[3]{%
				{\colorbox{#1}{\color{White}   \textsf{\textbf{#2}}} \textcolor{#1}{#3}}}

	}%
	{
		\newcommand{\ColorComment}[3]{}

	}%

\definecolor{cocoricolor}{RGB}{238, 130, 238}
\definecolor{michalcolor}{RGB}{255,127,80}
\definecolor{naphanncolor}{RGB}{112, 51, 173}\newcommand{\naphann}[1]{\ColorComment{naphanncolor}{whit3z}{#1}}
\definecolor{shigeyacolor}{RGB}{198,53,39}
\definecolor{rdvcolor}{rgb}{0,0.5,0}\newcommand{\rdv}[1]{\ColorComment{rdvcolor}{rdv}{#1}}
\definecolor{shotacolor}{rgb}{0,0,1}

\begin{document}

\date{}

\title{\Large \bf QuISP: a Quantum Internet Simulation Package}

\author{
{\rm Ryosuke Satoh}\\
Keio University
\and
{\rm Michal Hajdu\v{s}ek}\\
Keio University
\and
{\rm Naphan Benchasattabuse}\\
Keio University
\and
{\rm Shota Nagayama}\\
Mercari, Inc.
\and
{\rm Kentaro Teramoto}\\
Mercari, Inc.
\and
{\rm Takaaki Matsuo}\\
WIDE Project
\and
{\rm Sara Ayman Metwalli}\\
Keio University
\and
{\rm Takahiko Satoh}\\
Keio University
\and    
{\rm Shigeya Suzuki}\\
Keio University
\and
{\rm Rodney Van Meter}\\
Keio University
\and

} 

\maketitle
\input{chapters/0.Abstract}
\input{chapters/1.Introduction}
\input{chapters/2.Quantum}
\input{chapters/3.Design}
\input{chapters/4.Simulator}

\input{chapters/5.Performance}
\input{chapters/6.Conclusion}

\input{chapters/7.Acknowledgement}

\bibliographystyle{plain}
\bibliography{references}
\appendix
\input{chapters/A.Appendix}
\end{document}

%% file: chapters/0.Abstract.tex
\begin{abstract}
We present an event-driven simulation package called QuISP for large-scale quantum networks built on top of the OMNeT++ discrete event simulation framework.
Although the behavior of quantum networking devices have been revealed by recent research, it is still an open question how they will work in networks of a practical size. 
QuISP is designed to simulate large-scale quantum networks to investigate their behavior under realistic, noisy and heterogeneous configurations. The protocol architecture we propose enables studies of different choices for error management and other key decisions. Our confidence in the simulator is supported by comparing its output to analytic results for a small network. A key reason for simulation is to look for emergent behavior when large numbers of individually characterized devices are combined.
QuISP can handle thousands of qubits in dozens of nodes on a laptop computer, preparing for full Quantum Internet simulation. This simulator promotes the development of protocols for larger and more complex quantum networks.
\end{abstract}

%% file: chapters/1.Introduction.tex
\section{Introduction}
\label{sec:introduction}
The second quantum revolution has steadily been gaining momentum over the last two decades~\cite{dowling2003quantum}.
Quantum computers are at the forefront with private companies racing to build ever larger quantum devices demonstrating quantum supremacy~\cite{arute2019quantum, zhong2020quantum}.
Quantum networks~\cite{vanmeter2014quantum} promise to revolutionize the field of communication by exploiting the rules of quantum mechanics in order to bring enhanced, and in some cases completely new, functionality.
Some of the main promises of quantum networks include distribution of secret keys for secure communication where a malicious party is doomed to be discovered due to the fundamental laws of quantum mechanics~\cite{bennett2014quantum,ekert1991quantum}.
The possibility of distributed quantum computation where quantum computers are networked together in order to solve a difficult computational task in a cooperative fashion~\cite{cuomo2020towards} is being explored.
Blind quantum computation~\cite{broadbent2009universal} allows a client with limited quantum resources to delegate their computation to a powerful quantum server without revealing the input, the computation itself or its output.
Enhanced clock synchronization~\cite{jozsa2000quantum,ilo2018remote} promises to improve the accuracy of global navigation while distributed quantum states can be used to construct ultra-sensitive sensor networks~\cite{proctor2018multiparameter,zhuang2019physical}.
These applications are likely just the tip of the iceberg and novel uses of quantum networks will be discovered.

The fundamental resource behind these marvelous applications is entanglement~\cite{horodecki2007quantum}, the ability of spatially separated quantum states to be correlated more strongly than classical states.
The primary job of a quantum network is to distribute these entangled states between two or more parties.
The ultimate goal is to build a global Quantum Internet~\cite{kimble2008quantum,wehner2018quantum,satoh2021attacking} of entangled quantum devices, all running on heterogeneous hardware yet still being able to engage in reliable and efficient quantum communication.

Quantum networks are quickly becoming more than just a dream, with early quantum key distribution (QKD) networks having been implemented already~\cite{elliott2003quantum,peev2009theSECOQC,sasaki2011field,stucki2011long,dynes2019cambridge,chen2021integrated,joshi2020trusted}.
Currently, the race is on to demonstrate feasibility of repeater-based quantum networks, where entanglement over large distances is constructed recursively from shorter entangled links~\cite{briegel1998quantum,duan2001long}, with some of the basic hardware elements being demonstrated experimentally \cite{PRXQuantum.2.017002,jing2019entanglement,pompili2021realization}.

Besides the immediate technological challenges facing us when developing a global quantum network, we have to also deal with open questions for the future of the Quantum Internet.
This is where a quantum network simulator becomes a crucial research and design tool.
Some areas where a good simulator will be invaluable are:
(1) \emph{Protocol design}.
Particularly testing of detailed protocol design to validate correct operation and study of the interaction between classical and quantum portions of the network.
(2) \emph{Connection architecture and performance prediction}.
Proposed generations of quantum networks~\cite{muralidharan2016optimal} exhibit complex behavior making analytic prediction of their performance difficult with realistic parameters.
(3) \emph{Dynamic behavior}.
Important open questions are the stability of quantum networks as conditions change over time and the response of protocols to dynamically changing network topology.
(4) \emph{Emergent behavior}.
As the system increases in scale, new and unexpected behavior may emerge that current naive models of quantum networks do not take into account.
How do competing connections behave? In particular, are quantum networks subject to congestion collapse, or can short-distance connections starve long connections? Must we trade connection fidelity (quality) for performance? Naive models suggest that end-to-end, high fidelity connections are always possible, but it is still an open question whether this is true in a dynamic, global network.



Our approach to simulating large-scale quantum networks using our Quantum Internet Simulator builds on OMNeT++ \cite{varga2010omnet++}, a modular, component-based architecture simulation environment.
Its focus is on protocol design for complex, heterogeneous networks at large scale while keeping the physical layer as realistic as possible.
Following on from earlier simulations of single lines of repeaters~\cite{van-meter07:banded-repeater-ton,jiang2007oaq} and an early network simulator~\cite{aparicio2011master}, the past two years have seen a flurry of activity in the field of quantum network simulation, with the introduction of a number of simulators such as NetSquid \cite{coopmans2020netsquid}, SeQUeNCe \cite{wu21:sequence} and QuNetSim \cite{diadamo2020qunetsim}.
These simulators are limited by focusing on physically realistic simulation of a single, small network. QuISP is designed with internetworking in mind while maintaining full physical realism. Our long-term goal for the simulator is to be able to handle an internetwork with 100 networks of 100 nodes each, with each network running independent error management protocols, hardware parameters, and topology.

The manuscript is structured as follows. First, we provide basic concepts of quantum information processing and our new quantum state representation proposal towards to large  quantum simulation in Section~\ref{sec:background}. We then introduce the background design of quantum network architecture in Section~\ref{sec:design}. Our main proposal for our quantum networking simulator is described in Section~\ref{sec:using_simulator}. We give the detailed explanation of basic design principles and the current implementation of QuISP. Finally, we demonstrate experiments to show the correctness and performance of QuISP in Section~\ref{sec:performance} before concluding in Section~\ref{sec:conclusion}. Additional details on configuring the simulator and the mathematics of our new error basis simulation are in the appendices.

%% file: chapters/2.Quantum.tex
\section{Quantum}
\label{sec:background}
In this section, we give a brief overview of the basics of quantum information processing and discuss our approach to scalable simulation of quantum networks.

\begin{table*}[!t]
\centering
\caption{Classical representations of quantum states and their scaling}
\label{tab: simulation_scaling}
\begin{tabular}{|l|l|l|}
\hline
Representation & Description & Scaling \\ \hline

Full state vector  & \begin{tabular}[c]{@{}l@{}}every entry 000...0 to 111...1;\\ pure states (no error) only\end{tabular} & $O(2^n)$ \\ \hline

Dirac's bra-ket notation & \begin{tabular}[c]{@{}l@{}}sparse representation of state vector\\ (also used for variable names)\end{tabular} & $\ll O(2^n)$ but grows sharply to $O(2^n)$\\ \hline

\begin{tabular}[c]{@{}l@{}}von Neumann's \\ density matrix\end{tabular} & \begin{tabular}[c]{@{}l@{}}pure or mixed (error) states \\ (sparse representations also possible)\end{tabular}& $O(4^n)$ \\ \hline

Stabilizer~\cite{gottesman97:_thesis} & \begin{tabular}[c]{@{}l@{}}shorthand for specific states, specifies \\ constraints on states (eigenoperators)\end{tabular} & $O(n)$ to $O(n^2)$ \\ \hline

Tensor network~\cite{markov:contracting-siam,orus2014practical}  & tree-based, memoization-like~\cite{michie1968memo} & \begin{tabular}[c]{@{}l@{}}dependent on the amount of entanglement: \\up to 56 qubits, maybe 100 for very special\\ cases achievable~\cite{pednault:1710.05867,chen:1805.01450}\end{tabular} \\ \hline

\textbf{Error basis} & \begin{tabular}[c]{@{}l@{}}track only errors, not states\\ (useful for developing quantum error correction)\end{tabular}  & \begin{tabular}[c]{@{}l@{}}$O(n)$ for Pauli (symmetric) errors; \\ complex processing for others\end{tabular}\\ \hline
\end{tabular}
\end{table*}

\subsection{Quantum Information Processing}
Quantum networks encode information into quantum states \cite{nielsen2002quantum}.
The simplest and also the most commonly used quantum system is a \emph{qubit}.
It is comprised of two basis states, a $|0\rangle$ (pronounced ``ket 0'') and a $|1\rangle$ (pronounced ``ket 1'').
Unlike a classical bit that is either a 0 or a 1, the qubit can be in a \emph{superposition} of the two base states, written as $\alpha|0\rangle + \beta|1\rangle$, where $\alpha$ and $\beta$ are known as complex probability amplitudes.
They give the relative likelihood that the state of a qubit is a $|0\rangle$ or $|1\rangle$.
In particular, the state $|+\rangle=(|0\rangle+|1\rangle)/\sqrt{2}$ is an equal superposition of $|0\rangle$ and $|1\rangle$.
This superposition principle extends to multiple qubits as well.
In particular, qubits A and B can be in a superposition $\alpha|0\rangle_A|0\rangle_B + \beta|1\rangle_A|1\rangle_B$.
This is a rather special state because it is impossible to write it as a product of local qubit states of A and B.
Such states are known as \emph{entangled states} and are at the heart of most quantum technologies.
A particular entangled state used ubiquitously in quantum communication is when $\alpha=\beta=1/\sqrt{2}$, known as a \emph{Bell pair}.

The basic way to transform the state of a qubit is via \emph{Pauli operators} $X$, $Y$, and $Z$. Pauli $X$ is the quantum analogue of a bit-flip. It flips the state of a qubit $X|0\rangle = |1\rangle$ and vice versa. Pauli $Z$ does not have a classical counterpart as it introduces a \emph{phase}, namely $Z|1\rangle=e^{i\pi}|1\rangle=-|1\rangle$. This phase is important when the qubit is in a superposition state since $Z|+\rangle=(|0\rangle-|1\rangle)/\sqrt{2}$ is experimentally distinguishable from $|+\rangle$.

\emph{Quantum state tomography} \cite{altepeter2005photonic} is the process of characterizing the quantum state by measuring identical copies in different bases in order to reconstruct a statistical description of the state known as the \emph{density matrix}.
It is key to extracting useful information about the quantum state, particularly its \emph{fidelity} which is a measure of how close the actual output of a quantum protocol is to its desired one.

The state of a qubit can be transmitted in a quantum network using \emph{quantum teleportation} without physically sending the system encoding the qubit \cite{bennett1993teleportation}.
Consider a client wishing to communicate the state of qubit CQ$_{\text{1}}$ to a server.
They must share two qubits in a Bell pair, denoted by CQ$_{\text{2}}$ and SQ.
The client performs a \emph{Bell-state measurement} (BSM) on qubits CQ$_{\text{1}}$ and CQ$_{\text{2}}$, and sends the classical result of this measurement to the server who performs a conditional correction on its qubit SQ.
This ensures that the state of qubit SQ is the same as that of qubit CQ$_{\text{1}}$ was initially.
The entanglement between qubits CQ$_{\text{2}}$ and SQ is consumed during the measurement and therefore must be reestablished prior to any further quantum teleportations.
Note that entanglement does not imply any directionality so the server can transmit its qubit back to the client provided they share a Bell pair.

\begin{figure}[!t]
    \centering
    \includegraphics[width=0.9\linewidth]{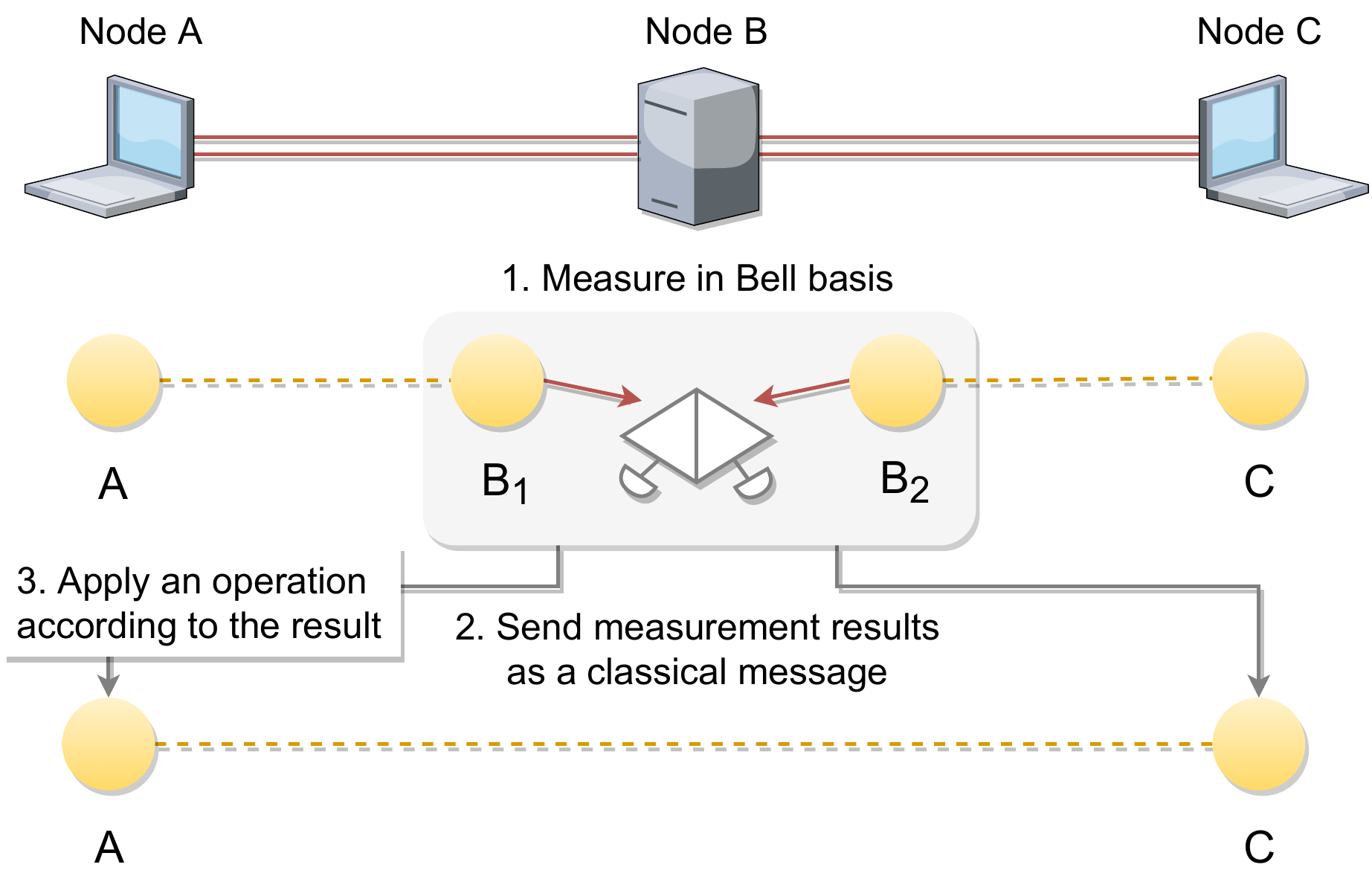}
    \caption{Once an intermediate node shares entangled pairs with its neighbors it measures the two shared qubits B$_{\text{1}}$ and B$_{\text{2}}$ in the Bell basis, resulting in qubit A becoming entangled with qubit C. }
    \label{fig: entanglement_swapping}
\end{figure}

Quantum teleportation can be extended to pairs of entangled qubits, in which case it is known as \emph{entanglement swapping} \cite{zukowski1993event}.
Consider two entangled pairs, denoted by A-B$_{\text{1}}$ and B$_{\text{2}}$-C as depicted in Figure~\ref{fig: entanglement_swapping}.
Measurement in the Bell basis of qubits B$_{\text{1}}$ and B$_{\text{2}}$ will result in qubits A and C becoming entangled.
Entanglement swapping is the basic building block of quantum repeater networks as it allows end-to-end entangled connection between distant nodes to be established.

Sharing entanglement between neighboring nodes of a quantum network is not a deterministic process due to noise and losses in the fiber.
Even once the entanglement is established its quality deteriorates over time due to decoherence of quantum memories.
One of the jobs of a quantum network is to manage these errors via \emph{entanglement purification} \cite{bennett1996concentrating}.
In this protocol, two imperfect entangled pairs can be converted probabilistically into a single entangled pair of higher fidelity.

\subsection{Simulating in the Error Basis}
\label{sec: error_simulation}

Quantum systems are exceptionally fragile and susceptible to errors due to interaction with the environment.
Furthermore, distant nodes in a quantum network communicate via exchange of single photons, meaning photon loss in the optical fiber becomes a major source of errors.
It is therefore crucial that a quantum network simulator models all physically relevant sources of errors accurately.

Calculating the state of a quantum system by hand or classical computer can be done using one of several representations, with differing purposes and scaling, as shown in Table~\ref{tab: simulation_scaling}. Many important states can be written down using shorthand methods such as the sparse Dirac bra-ket notation or \emph{stabilizers}~\cite{gottesman97:_thesis}, but complete representation of an arbitrary state may require up to $2^n$ complex numbers when the state is \emph{pure}, or error free. Using special techniques on classical supercomputers, computations of special cases involving up to 56~\cite{pednault:1710.05867} or 81~\cite{chen:1805.01450} qubits have been simulated.

Traditional tracking of the evolution of a \emph{mixed}, or noisy, quantum state composed of $n$ qubits is substantially harder.
The \emph{density matrix} can represent any quantum state after any computation or noise process, but when fully populated, requires $2^n\times 2^n$ complex numbers. 
In the context of quantum networks, full density matrix simulation is tractable for 1G networks, where the simulation must track many independent quantum states, but each state is composed of at most four qubits. This approach has been used in numerous simulations, especially those focused on tuning low-level hardware control parameters~\cite{ladd06:_hybrid_cqed,pathumsoot2020modeling}.
However, this approach quickly becomes intractable when moving to 2G and 3G networks, where logical Bell pairs are encoded into a large multi-qubit state for quantum error correction~\cite{muralidharan2016optimal}. Simulating only two 2G links using a 7-qubit error correcting code on each link~\cite{PhysRevA.79.032325} requires $16 \times 2^{56}$ bytes, about one exabyte. Simulating a 100 hop long connection could take a mind-boggling $3.2\times 10^{3011}$ bytes of memory. Thus, it is imperative to separate simulators based on their purpose; our goal is to study the architecture and protocols of networks, rather than the physics of devices or complete quantum algorithms.

QuISP works on the premise that the desired quantum state is known and we only have to track deviations from this ideal state in the form of errors that are affecting it.
This approach is adapted from quantum error correction~\cite{devitt2013quantum, nagayama21:e2eep} and represents a novel way of simulating quantum networks.
Deviations from the ideal state can be tracked efficiently leading to scalable simulation of truly global quantum networks beyond 1G for the first time.
This scalability is achieved by exploiting the fact that quantum networks only need to apply a fixed number of operations and the major sources of errors can be discretized into a small set.
Our approach is tailor-made for quantum network simulation and is not suitable for simulation of a universal quantum computer.

The simulator tracks Pauli $X$, $Y$, and $Z$ errors discussed in Section~\ref{sec:background}.
It also tracks \emph{relaxation} and \emph{excitation errors}, denoted by $R$ and $E$, respectively.
Relaxation captures the process of energy loss due to environmental noise by transforming any initial state to $|0\rangle$, while excitation captures the opposite process where any arbitrary initial state is transformed to $|1\rangle$.
Finally, the simulator tracks \emph{photon loss} $L$ as well.
This list does not form an exhaustive set of all possible possible errors and may be expanded or shrunk depending on the particular physical scenario under consideration.

The quantum state at time $t$ is represented by a $m+1$-element \emph{error probability vector} $\vec{\pi}(t)$, where $m$ is the number of errors that we are tracking.
Each element $\pi_j$ represents the probability that the quantum state has been affected by a particular error $j$ at time $t$,
\begin{equation}
    \vec{\pi}(t) = \begin{pmatrix} \pi_I & \pi_X & \pi_Y & \pi_Z & \pi_R & \pi_E & \pi_L \end{pmatrix}.
    \label{eq:pi_q7}
\end{equation}
The error probability vector is normalized, meaning $\sum_j \pi_j=1$.

The evolution of the error probability vector is given by the \emph{transition rate matrix} $Q$.
It is a right stochastic matrix with elements $Q_{ij}$ representing the probability per unit time of transitioning from error state $\pi_i$ to error state $\pi_j$, and they satisfy the normalization condition $\sum_{j} Q_{ij} = 1$.
The error probability vector $\vec{\pi}(t)$ a time $t$ is given by
\begin{equation}
    \vec{\pi}(t) = \vec{\pi}(t-1) Q = \vec{\pi}(0) Q^t.
    \label{eq:pi_evolution}
\end{equation}
The transition rate matrix $Q$ is independent of time and the error probability vector $\vec{\pi}(t)$ depends only on $\vec{\pi}(t-1)$, meaning the evolution of the qubit is modelled as a Markov process.
Detailed discussion of the transition matrix can be found in Appendix~\ref{sec:transition_matrix}.

The simulator samples the error probability vector $\vec{\pi}(t)$ prior to an operation on the qubit and applies the sampled error type%
\footnote{The current version of the simulator samples the error probability vector before a measurement or purification is performed. The next version will extend the sampling to entanglement swapping as well.}.
The simulation must be repeated to gather statistics about the real state of the qubit and gain information about the qubit's fidelity.

Extension to multi-qubit systems is straightforward.
The state of $N$ qubits in QuISP is described by an $N(m+1)$-element error probability vector where the first $m+1$ entries describe qubit 1, second group of $m+1$ describe qubit 2 and so on.
The full transition rate matrix for $N$ qubits is given by a block-diagonal matrix composed of $N$ single-qubit transition rate matrices.


%% file: chapters/3.Design.tex
\section{Network Design}
\label{sec:design}


In this section, we discuss the basic network design that QuISP assumes.

\subsection{Quantum Network Architectures}

\begin{figure}[!t]
    \centering
    \includegraphics[width=0.8\linewidth]{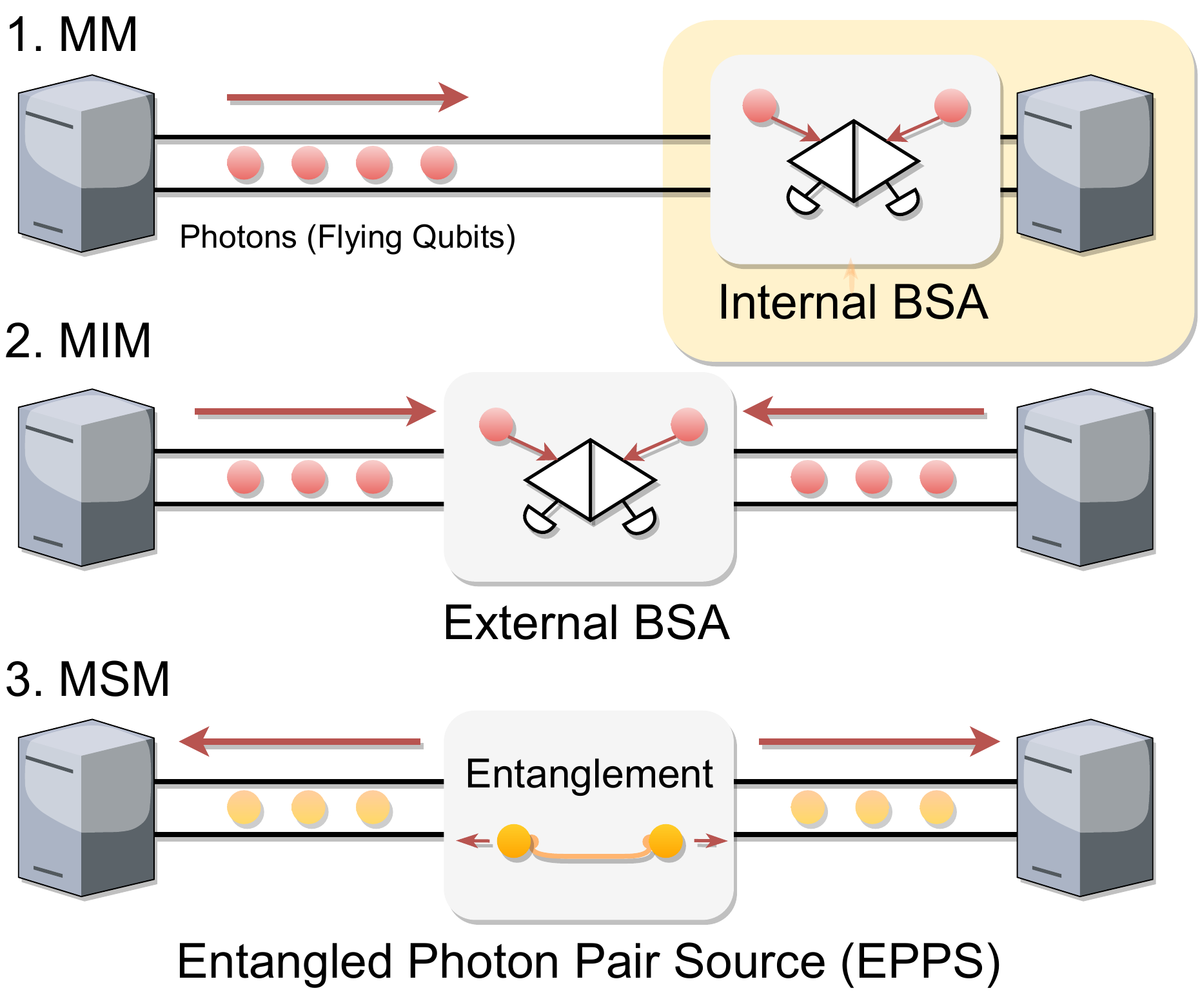}
    \caption{Three quantum link architectures. MM and MIM differ mainly by the position of the BSA while MSM replaces the BSA in the middle with an EPPS.}
    \label{fig:link_architecture}
\end{figure}


There is currently no consensus on the best overarching network architecture for quantum networking, though the key principles are coming into view~\cite{I-D.irtf-qirg-principles, van-meter21:aqia} and some protocol elements have been proposed~\cite{dahlberg2019link,kozlowski00:qnp,matsuo2019quantum}.
Supporting further research in this area is the primary purpose of QuISP. However, it is becoming clear that some basic hardware and software components will most likely be shared between future candidate architectures.
We give a brief outline of these components in this subsection.

\textbf{Hardware architecture}:
There exist a number of potential candidate physical systems that are suitable for encoding qubits, broadly divided into two categories.
\emph{Stationary qubits} or \emph{matter qubits} are envisioned to store and process information at the nodes of a quantum network, acting as the hosts.
Candidate physical systems include nitrogen-vacancy centers in diamond~\cite{pompili2021realization}, trapped ions~\cite{duan2010colloquium}, atomic ensembles~\cite{yu2020entanglement}  and superconducting qubits~\cite{mirhosseini2020superconducting}.


Inter-node quantum communication is achieved by using \emph{flying qubits} encoded onto single photons travelling through optical fibers.
We refer to these fibers as \emph{quantum links}.
Photons are ideal information carriers as they do not interact strongly with their environment and they travel at very high speeds.
Using flying qubits, it is possible to distribute entangled Bell pairs between two distant nodes of a quantum network.
This can be achieved using one of the three existing quantum link architectures~\cite{jones2016design} depicted in Fig.~\ref{fig:link_architecture}.
\emph{Memory-Memory} (MM) link connects two quantum nodes directly where either node is equipped with a \emph{Bell-state analyzer (BSA)}, an optical device that performs a Bell-state measurement on two incoming photons.
\emph{Memory-Interfere-Memory} (MIM) link places the BSA in the middle of the quantum link.
\emph{Memory-Source-Memory} (MSM) link replaces the the BSA in the middle of the quantum link with a source of entangled photonic pair states (EPPS).
Despite the architectures appearing fairly similar, they differ significantly in their performance as well as the technological maturity required to implement them.

Since quantum communication operates at the single-photon level, attenuation becomes a major source of error.
Unlike classical bits, qubits cannot be copied and resent owing to the \emph{no-cloning theorem}~\cite{wootters1982single}, a fundamental property of quantum mechanics forbidding to deterministically copy arbitrary states of quantum systems.
Amplification at the level of single photons becomes ineffective as well~\cite{chia2019phase,chia2020phase}.
This limits the practical length of quantum links to mere tens of kilometers.

\begin{figure}[!t]
    \centering
    \includegraphics[width=\linewidth]{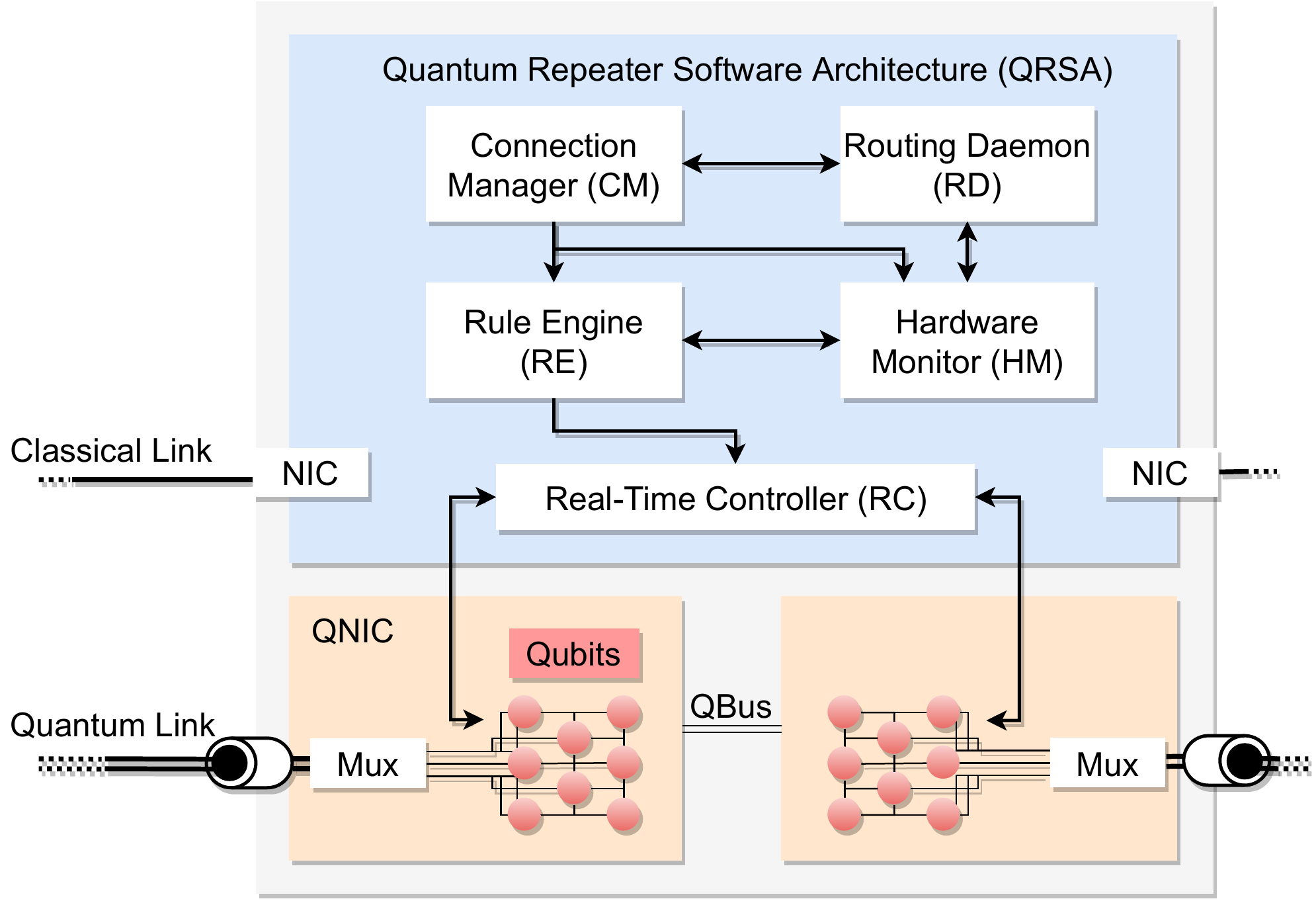}
    \caption{Our target quantum repeater architecture. The top blue section denotes QRSA composed of five distinct software components discussed in the main text. The arrows out of each component represent directions of messages. The bottom orange sections are QNICs which contains multiple stationary qubits to hold quantum information, and manipulate these qubits to extend entanglement via entanglement swapping. 
    }
    \label{fig: repeater}
\end{figure}

In order to get around this problem, a new type of node was introduce known as a \emph{quantum repeater}~\cite{briegel1998quantum,duan2001long}.
One of the jobs of a quantum repeater is to share Bell pairs with its neighboring nodes and implement entanglement swapping in order to create a Bell pair between these nodes.
In this way, the no-cloning theorem can be sidestepped and photon loss mitigated, resulting in the possibility of establishing end-to-end Bell pairs between arbitrarily separated quantum hosts.
Figure~\ref{fig: repeater} depicts the hardware and software components of a quantum repeater.


A particularly important component of the quantum repeater is the \emph{Quantum Network Interface Card} (QNIC).
The QNIC is the quantum analogue of a classical NIC with one major difference being that a QNIC is able to apply quantum operations to the store quantum information, making it a quantum computer with limited capabilities.
In particular, the QNIC is capable of applying single-and two-qubit gates as well as single- and two-qubit measurements.




\textbf{Software architecture}
Classical software running on a quantum repeater will play a crucial role when designing efficient repeater-based quantum networks.
Our proposed software architecture, \emph{Quantum Repeater Software Architecture} (QRSA), as shown in Figure~\ref{fig: repeater}, consists of five software components.


\begin{itemize}
\item \emph{Connection Manager} (CM):
 CM manages the connection from the Initiator to the Responder. Once a connection setup request is initiated at an Initiator, it is passed to a Responder through a specific path. At this point, intermediate nodes provide the required information, such as QNIC interface information. The most important task for the Connection Manager is to generate \textit{RuleSets}, which we discuss in Section~\ref{sec: RuleSet}.

\item \emph{Hardware Monitor} (HM):
HM is responsible for monitoring quantum links between the neighboring network nodes. In quantum networking, the quality of links is critical to the final quality of the end-to-end Bell pair. The HM collects information about fidelity and generation rate that is used by RD and CM.

\item \emph{Rule Engine} (RE):
The main responsibility of the RE is executing RuleSets issued by the CM.
To achieve this, the RE constantly monitors the quantum resources available and manages these resources.
The results of executed actions are reported back to the RE, and are also distributed to partner nodes where appropriate. 
RE updates the state of qubits based on incoming messages from itself and other nodes.

\item \emph{Real-Time Controller} (RC):
RC is in charge of initializing physical qubits and coordinating their photon emissions for the purpose of entanglement distribution.
The RC selects which qubits are scheduled to emit photons and at what time.
After the qubits no longer take part in entanglement distribution, the RC  reinitializes them. RC is device drivers and lower-level software with a hard real time component, interfacing directly to the hardware.


\item \emph{Routing Daemon} (RD):
RD's responsibility is to create and maintain the routing table for the quantum interfaces. It exchanges information with RDs in neighboring nodes in accordance with a standardized routing protocol. It conveys the information about route and QNIC identifiers required to reach other end node (destination) to other components of the QRSA.

\end{itemize}

These components communicate as needed to convey when to start operations and the current status of devices. 

An end node has almost the same functionality as a repeater, but it also has an Application component responsible for performing end-to-end applications.

\subsection{RuleSet Protocol}
\label{sec: RuleSet}
In order to distribute end-to-end entanglement, both end nodes and quantum repeaters must know what actions to perform, when to execute them, and what other nodes are taking part in the process if the actions need to be coordinated.
For example, the repeater must know the nodes that it shares Bell pairs with when executing entanglement swapping since the results of the procedure must be shared with those nodes.

\begin{figure}[]
    \centering
    \includegraphics[width=\linewidth]{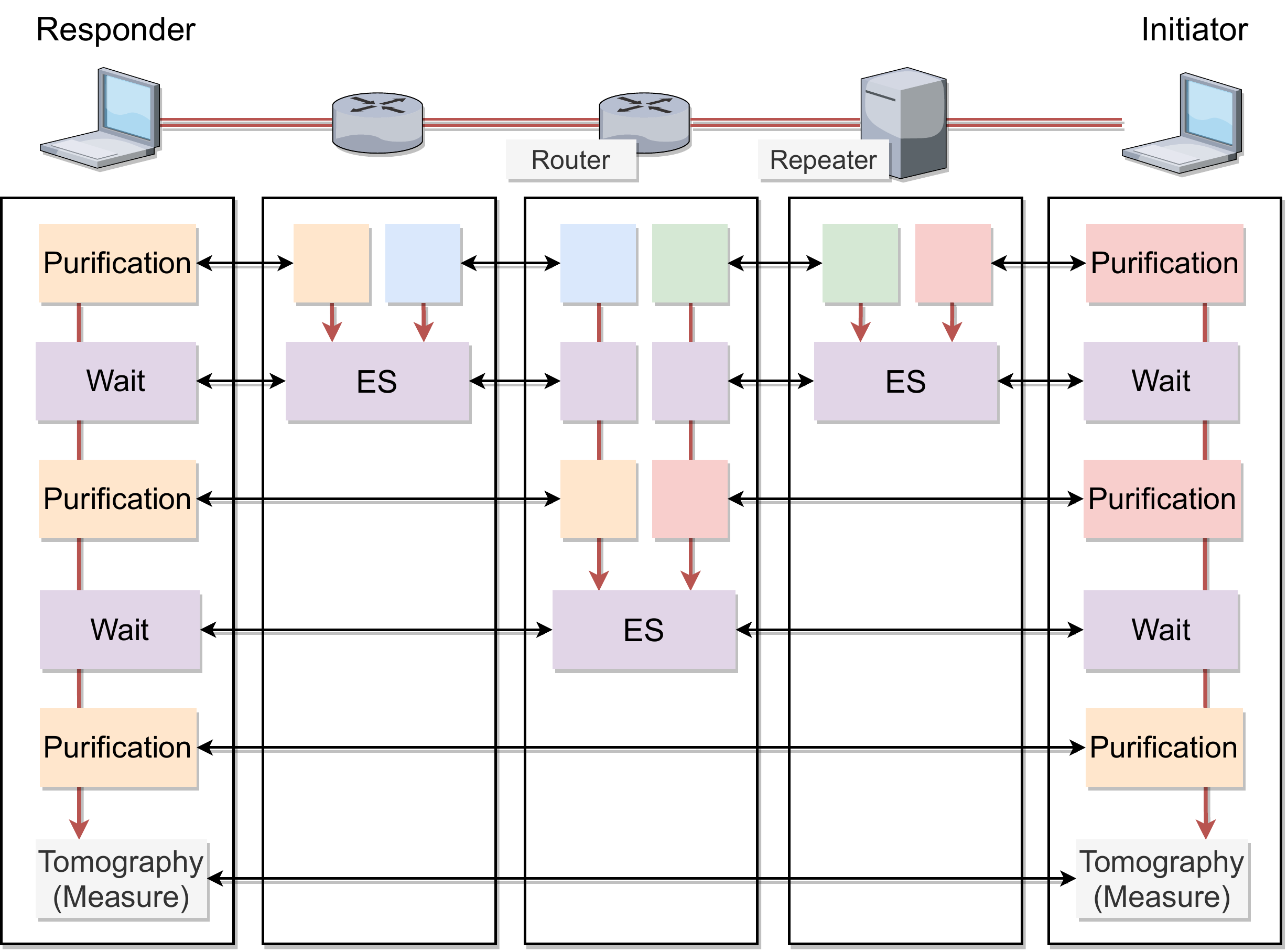}
    \caption{Example of RuleSets. Each node in the path has one RuleSet for the connection. \textit{Rules} are executed from top to bottom while communicating to the proper operation partners. Horizontal arrows represent the partners that are coordinating actions and vertical arrows represent the order of execution. ES is entanglement swapping.}
    \label{fig: ex_ruleset}
\end{figure}

To this end, the \textit{RuleSet} protocol was proposed in~\cite{matsuo2019quantum}, which QuISP supports.
The goal of this protocol is decentralized, autonomous but coordinated actions of the quantum repeaters with minimal classical inter-node communication. Figure~\ref{fig: ex_ruleset} is an example of RuleSet structure. The RuleSet is a collection of \textit{Rules} such purification and entanglement swapping. These RuleSets are built in the connection setup phase discussed in Section~\ref{sec: connection_setup}, and executed in a specified order. After being acted upon, the Bell pairs belonging to a particular Rule are passed to the next in sequence.

\begin{figure}[!t]
    \centering
    \includegraphics[width=\linewidth]{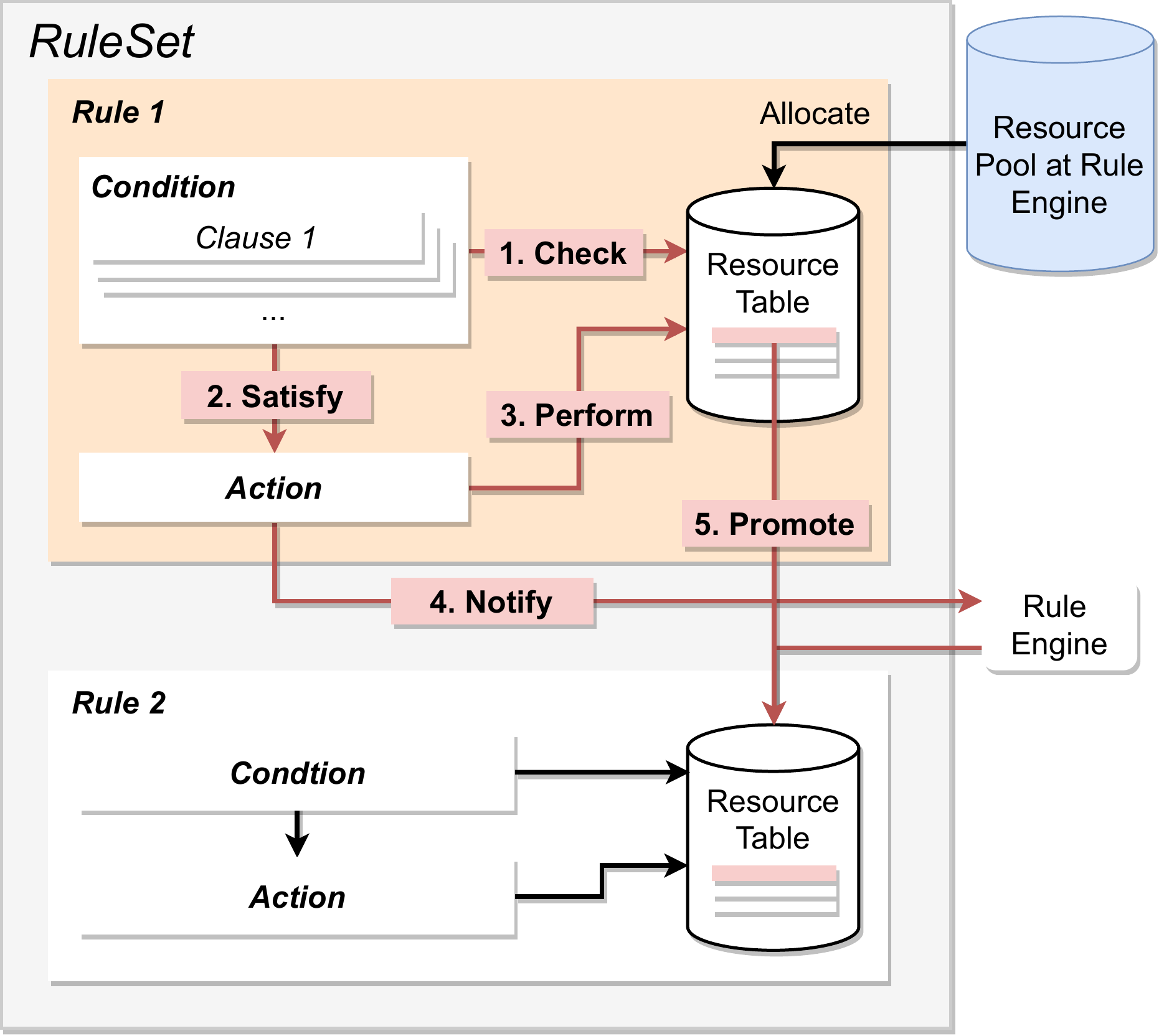}
    \caption{RuleSet execution. 1. Condition clauses checked one by one. 2. If all condition clauses are met, go to step 3, otherwise goes back to step 1 and wait for next allocation of resources. 3. Rules start performing actions. 4. The action notifies the result to the RuleEngine. 5. RuleEngine promotes the resource from one Rule to the next.}
    \label{fig: RuleSet}
\end{figure}

Figure~\ref{fig: RuleSet} describes the details of RuleSet and Rule. Every Rule has a \textit{Condition} and corresponding \textit{Action}.
The Actions are executed upon satisfaction of local conditions, usually relating to the number and quality of available quantum resources (Bell pairs). 
In quantum networking, shared Bell pairs must be managed by each node in a coordinated fashion and appropriately structured RuleSets provide this required consistency in terms of quantum operations.

Condition Clauses are composed of single or multiple conditions to be met before the Action is executed. For example, if node A requires two entangled states with node B to perform one action, A must track the number of shared entangled states with B. In such a situation, the Condition used is the \textit{Enough Resource Clause},  which is satisfied when the number of total entangled pairs shared with the proper partner is larger than a threshold (In this case, the threshold is two). Other than \textit{Enough Resource Clause}, there are several clauses supported in this simulator. 

An Action Clause is a set of operations including resource assignment changes, qubit manipulation, and classical message transfer. Once Condition Clauses are met, the corresponding action is immediately executed.  For example, \textit{Swap} refers to the resource table that belonging to the Rule and recognizes the corresponding qubits. Then, this action chooses one state entangled to its left and one entangled to its right and applies Bell state measurement. Informing the partners of the result of the Bell state measurement is one responsibility of the action.   

\subsection{Connection Setup}
\label{sec: connection_setup}
\begin{figure}[!t]
    \centering
    \includegraphics[width=\linewidth]{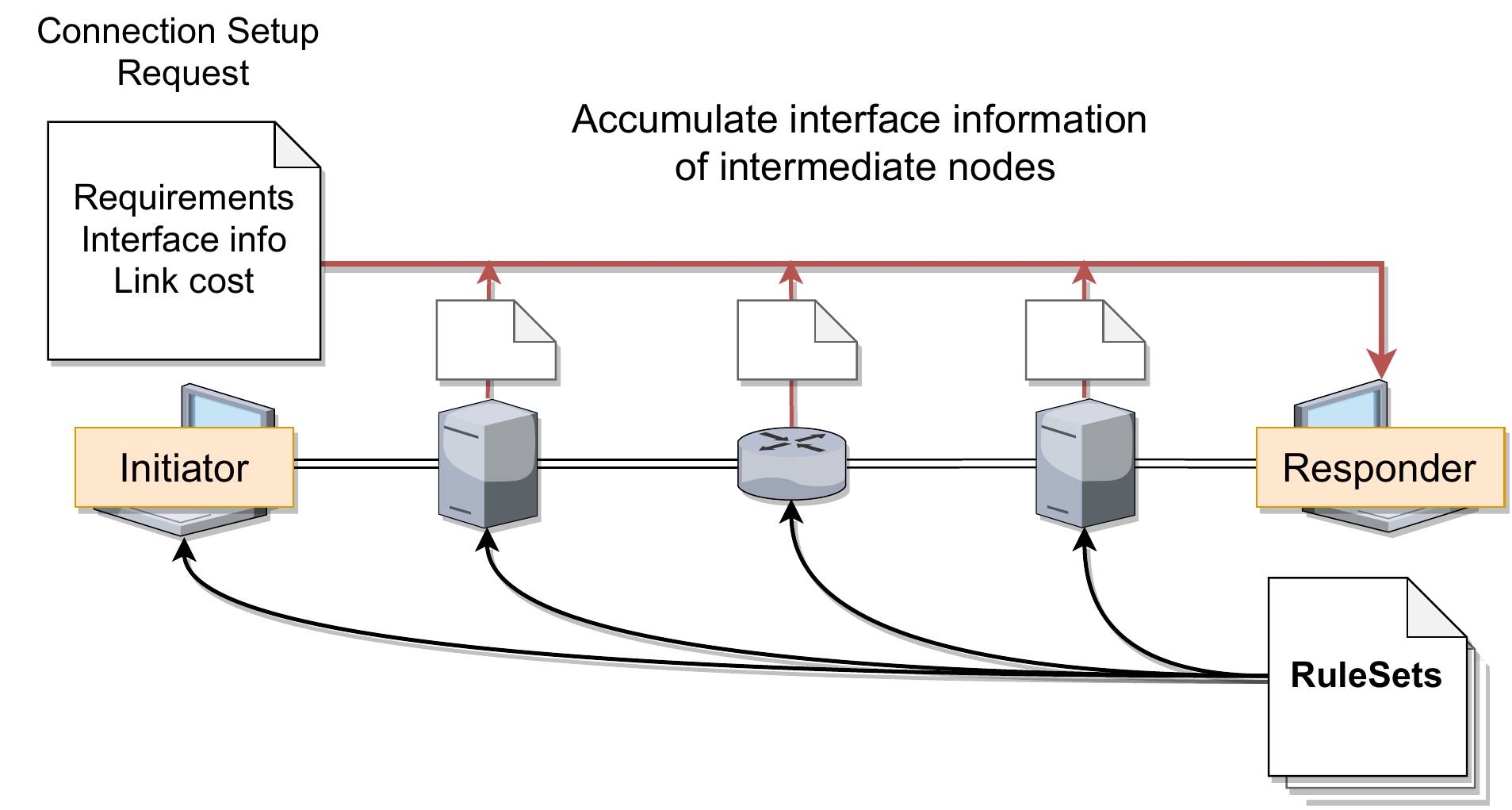}
    \caption{Connection setup process to establish agreements between initiator and responder}
    \label{fig: connection_setup}
\end{figure}

The connection setup is the step requires to gather all the information to create RuleSets to be executed for nodes that will be participating in the end-to-end Bell pair generation. The connection setup process used in QuISP is adapted from protocol outlined by Van Meter and Matsuo~\cite{I-D.van-meter-qirg-quantum-connection-setup}. Figure~\ref{fig: connection_setup} shows the procedure of the connection setup. It involves a two pass process, gathering the link information along the path starting from the node that tries to establish the connection (Initiator) and planning the RuleSet to be distributed among the nodes along the path at the other half of the connection (Responder). 

The first part, at the Initiator node, it receives the requirement for the connection from application level, like the quality of the connection (fidelity of the end-to-end Bell pair) and the number of Bell pairs. 
In this outbound pass, every node along the path will include their link characteristics into the message, reserve the QNIC, and relay this connection setup message to the next hop. 
If the QNIC cannot be reserved because it is already in use for another connection, the node will reject the request and the connection setup reject message will be sent to all the previous nodes along the path. 

When the connection setup message arrives at the Responder node, the Responder's job is to plan out how each node should execute their share of work in order for the end-to-end Bell pair creation to succeed. 
After planning out and creating RuleSets for all nodes, the Responder sends \emph{Connection Setup Response} with the RuleSets back to all nodes along the path.





%% file: chapters/4.Simulator.tex
\section{The Simulator}
\label{sec:using_simulator}



In the previous section, we explained the network design underlying QuISP. Here we introduce its implementation and use, with an example simulation.


\subsection{Goals and requirements}
\label{sec: goals}
The long-term goals for developing QuISP are supporting 1G, 2G, and 3G quantum networking (generations classified by the type of error handling) and internetworking in a scalable manner to establish highly reliable protocols for each generation. In the higher generations of quantum networking, thousands of qubits will work together when creating a single end-to-end logical Bell pair.
Furthermore, this simulator will allow us to replicate the dynamic behavior of quantum networks and internetworks with complex network topologies, including the concepts of heterogeneity and network boundaries. 

As noted in the introduction, our long-term goal is 100 networks of 100 nodes each. With inter-repeater spacing as low as 10km for some hardware architectures, long chains of repeaters between routers will be necessary to achieve distance. With low initial link performance, the router-level topology will have to be rich in order to provide adequate connectivity and resource availability. We expect this scale of simulation to aid in assessing an internetwork with such real-world constraints.



\subsection{Basic Design Principles}
We now highlight the main principles that we used as a guide in designing QuISP.

\textbf{Realism}.
The simulation should provide accurate information about the physical states that are being distributed in the quantum network.
This in turn provides accurate information about the fidelity of the distributed states.
Emphasis in the design of QuISP has been placed on realistic noise models.
Qubits stored in quantum memories will undergo decoherence processes that wash away their quantum properties and result in deteriorating fidelity.
Photon loss in fiber is a major source of error and therefore must be accounted for in the simulator. Early simulation work assumed $2^n$ hops, all of the same length and fidelity~\cite{jiang2007oaq,van-meter07:banded-repeater-ton}, but that is not realistic.

The implementation of the physical layer simulation must be kept cleanly separated from the implementation of software for the nodes themselves. \emph{The router software can know only what it would know in the real world}: information it can learn from classical measurement outputs from the quantum hardware and from exchanging messages with other nodes.

\textbf{Scalability}.
One of the main goals of this simulator is the study of large-scale quantum networks and their emergent complex behavior. Scalability concerns come in two flavors: the number of qubits involved in a single quantum state, and the number of nodes, links and connections in the whole network.
General multi-qubit states are difficult to simulate classically.
Our novel approach of working in the error basis discussed in Section~\ref{sec: error_simulation} allows us to simulate quantum networks at the cost of only a few classical bits per qubit,
and makes QuISP suitable for simulation of truly large-scale quantum networks of any generation.

\textbf{Flexibility}.
This simulator has been designed to offer the user a great deal of customizability.
Currently, there exist a number of physical systems that are considered to be promising candidates for implementing the physical layer of a quantum network.
However, all these candidate systems differ substantially in terms of parameters such as memory coherence times and operation clock rates.
QuISP allows the user to customize the error types that affect the physical layers as well as the individual parameters corresponding to error rates.
This ensures that QuISP is capable of accurately simulating any current and future quantum hardware.
Furthermore, the topology of the quantum network can be defined easily and this simulator comes with a set of predefined number of qubits per node and topologies as well. 

\subsection{Event Simulation with OMNeT++}
\label{sec: omnetpp}
OMNeT++ is a C++-based discrete event simulator primarily for classical networking simulations~\cite{varga2001discrete, omnetpp_hp, omnetpp_repo, varga2010omnet++}.  
The OMNeT++ model consists of \emph{modules} that communicate via \emph{messages}. The first batch of messages is created during module initialization. Their destination can be any other modules connected to the sender module or to self. Upon receiving a message, which constitutes an \emph{event}, the module processes the message and usually creates another batch of messages to be sent. The only proper way to trigger execution of a function in a certain module is by receiving messages. It is assumed that the processing of events takes zero time and that nothing of significance occurs between two events. The simulation ends when there are no messages left to be processed or the simulation reaches the time limit set by the user. The simulation time is tracked by the timestamp of events, which is determined by the time a message is scheduled to be sent.

Modules and messages in OMNeT++ are provided by the \verb|cSimpleModule| and \verb|cMessage| class, accordingly. \verb|cSimpleModule| is the active component in OMNeT++ and its functionality can be extended by adding C++ code or by composing multiple simple modules into a compound module.

Furthermore, QuISP also provides a graphical user interface from the OMNeT++ IDE to manipulate the network topology stored in a NED file, which defines network devices in very structured manner. An example of a NED file is shown in Appendix~\ref{sec: ned_definition}. The simulator also allows the user to change the initial simulation parameters or variables on launch using an INI configuration file.
It is also possible to visualize many stats, such as the changes of parameters during the simulation. These functions allow us to debug and check if the simulation works properly and to make intuitive performance measurements.


\subsection{Basic Implementation}
\label{sec: background_implementation}

\begin{figure}[!t]
    \centering
    \includegraphics[width=0.8\linewidth]{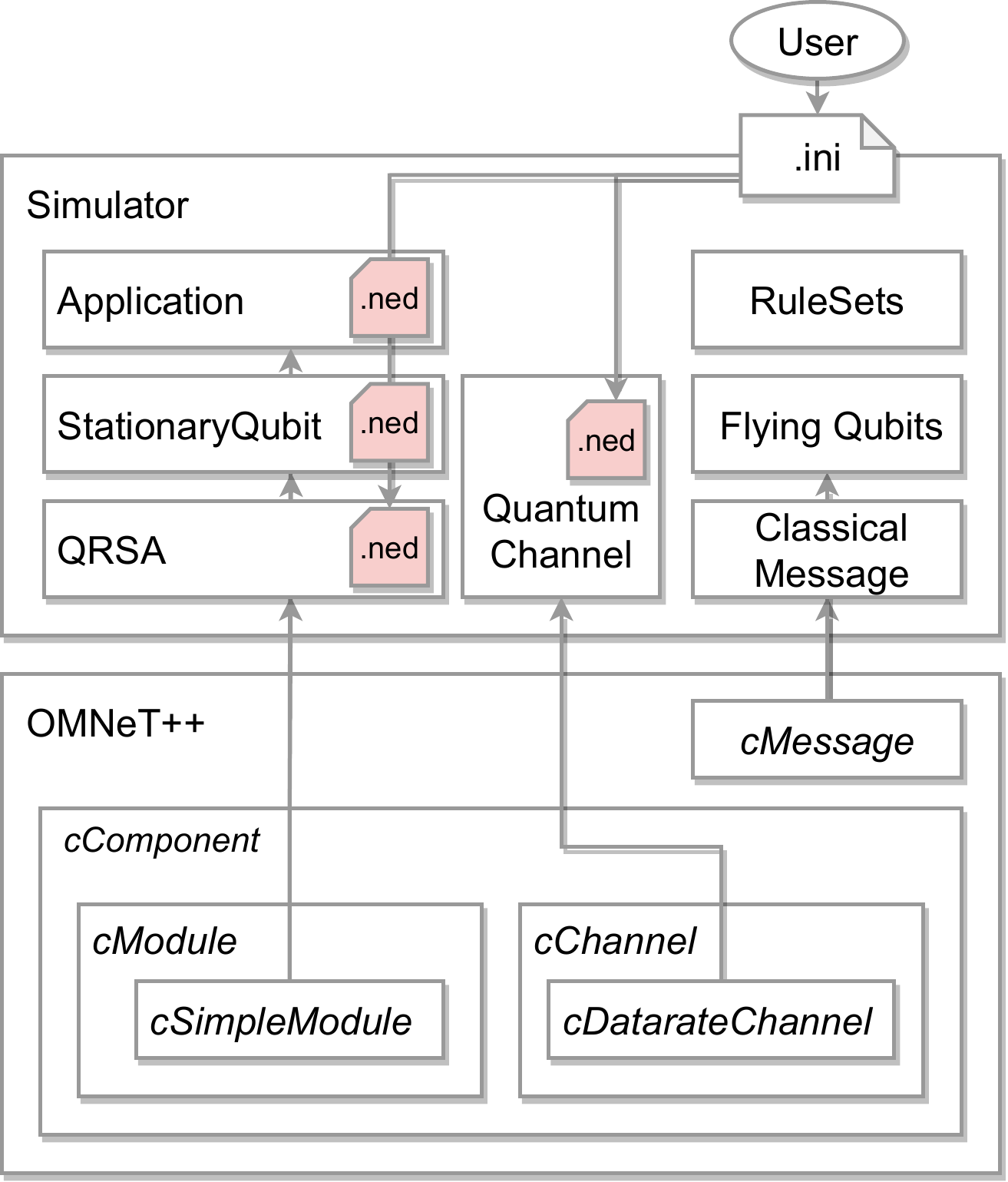}
    \caption{The implementation of QuISP. Users can provide simulation configurations as an INI file that can change variables of defined modules. Basic software components and channels are implemented as subclasses of \texttt{cSimpleModule} and \texttt{cDatarateChannel}. Messages and flying qubits emitted from those components are defined in \textit{.msg} files, which are converted to program source when they are compiled by OMNeT++.}
    \label{fig: background_imple}
\end{figure}

A concise diagram outlining our implementation is shown in Figure~\ref{fig: background_imple}.
Software modules (QRSA, Application, etc.) are built on top of \verb|cSimpleModule|, which supports standard message handlers. Stationary Qubit is also a subclass of \verb|cSimpleModule|, but it holds its quantum properties as its variables and behaves as quantum memory. 
Those software components can emit registered classical packets such as \verb|ConnectionSetupRequest| written in \verb|cMessage|.
Flying qubits can be thought of as an emitted message from Stationary Qubit. Thus a flying qubit is implemented as an extension of a classical packet passed through quantum channels. 

A quantum channel as a carrier of the flying qubit is implemented as a subclass of  \verb|cDatarateChannel|. The errors affecting flying qubits are calculated based on the distance between two nodes. 



\subsection{Supported Functionality}
\label{sec: functionality}
QuISP currently supports end-to-end tomography-based quality analysis of connections as the application. The output of the simulation is the characteristics of the path used for each connection, explained in detail in Section~\ref{sec:performance}.

The user can define complex topologies easily. Almost every factor that plays a role in the link characteristics such as node placement, distance between nodes, link types, number of buffer qubits in each interface,  channel error and loss rate, and the efficiency of most components that contribute to 1G and 2G technologies can be set individually or from defaults.

The simulator also supports options for selecting how the end-to-end Bell pair is generated. The order of entanglement swapping can be chosen between the usual binary tree like structure~\cite{briegel1998quantum} or swap-once-ready
at each node. The user can also tinker with the number of rounds of purification performed before entanglement swapping takes place and choose one of the four types of purification methods~\cite{briegel1998quantum, fujii2009entanglement} we currently provide. 

Since most of the groundwork for the RuleSet-based protocol has been implemented, curious users
can modify the simulator code, write their own entanglement swapping policy or purification method, and experiment with how to achieve higher quality connections.

\subsection{Configurable Parameters}
\label{sec: inifile}

As we mentioned in the previous section, users can make their decisions on the error rate, the number of purifications, the way of entanglement swapping. 
Currently, there are more than 50 types of parameters that can be given through the INI file. The following parameters are most likely to be configured by users.

\begin{itemize}
\item \textit{Network}. The name of the network written in the \textit{NED} language, specifying the number and types of nodes and the network topology.
\item \textit{Channel errors}. The errors on a quantum channel. This specifies how much and what types of errors in a particular distance. Currently, all Pauli errors and photon loss errors are supported, and users can specify the ratio of each error.

\item \textit{Memory errors}. In addition to Pauli errors, we support relaxation and excitation errors. As the simulation time passes, this error accumulates on each qubit. 

\item \textit{Gate errors}. In a realistic situation where we assume that devices are imperfect, there are always small operational errors in quantum gates. We can handle errors on basic gate sets (Pauli gates, Hadamard, Controlled-X). 

\item \textit{Measurement errors}. Quantum memories must be measured to get internal information which affects the quantum state itself. Measurement errors are composed of Pauli errors.

\item \textit{BSA errors}. It is possible to have different errors for internal and external BSAs. Tunable parameters are photon loss, photon detection rate, and dark count probability.

\item \textit{Photon emission success probability}. The probability that a quantum memory successfully emits a photon which is then captured by the fiber. 

\item \textit{Traffic pattern}. A configurable parameter for traffic generation pattern. Currently, this simulator supports single traffic generation from one node to the other random node and multiple traffic generation from all end nodes to randomly selected nodes.

\item \textit{Measurement count}. The number of measurement in tomography. When the required precision of tomography is higher, the number of measurement needs to be larger.
\end{itemize}

\subsection{Simulation Output}
\label{sec: performance}
The simulator returns the performance statistics at the end of simulation when the target configuration contains tomography. The output files contain the following information.
\begin{itemize}
\item \textit{Fidelity}. Calculated by the software tomography process, this indicates the quality of Bell pairs between two end nodes.
\item \textit{Bellpair per sec}. The number of Bell pairs generated in one second (in simulation time).
\item \textit{Tomography time}. The time until the entire tomography process finishes. 
\item \textit{Tomography measurements}. The number of measurements of tomography.
\end{itemize}
Besides these properties, the output statistics include more link information. An example output is given in Listing~\ref{lst: result} in Section~\ref{sec: example}.

\subsection{Example Simulation}
\label{sec: example}
\begin{figure}[]
    \centering
    \includegraphics[width=\linewidth]{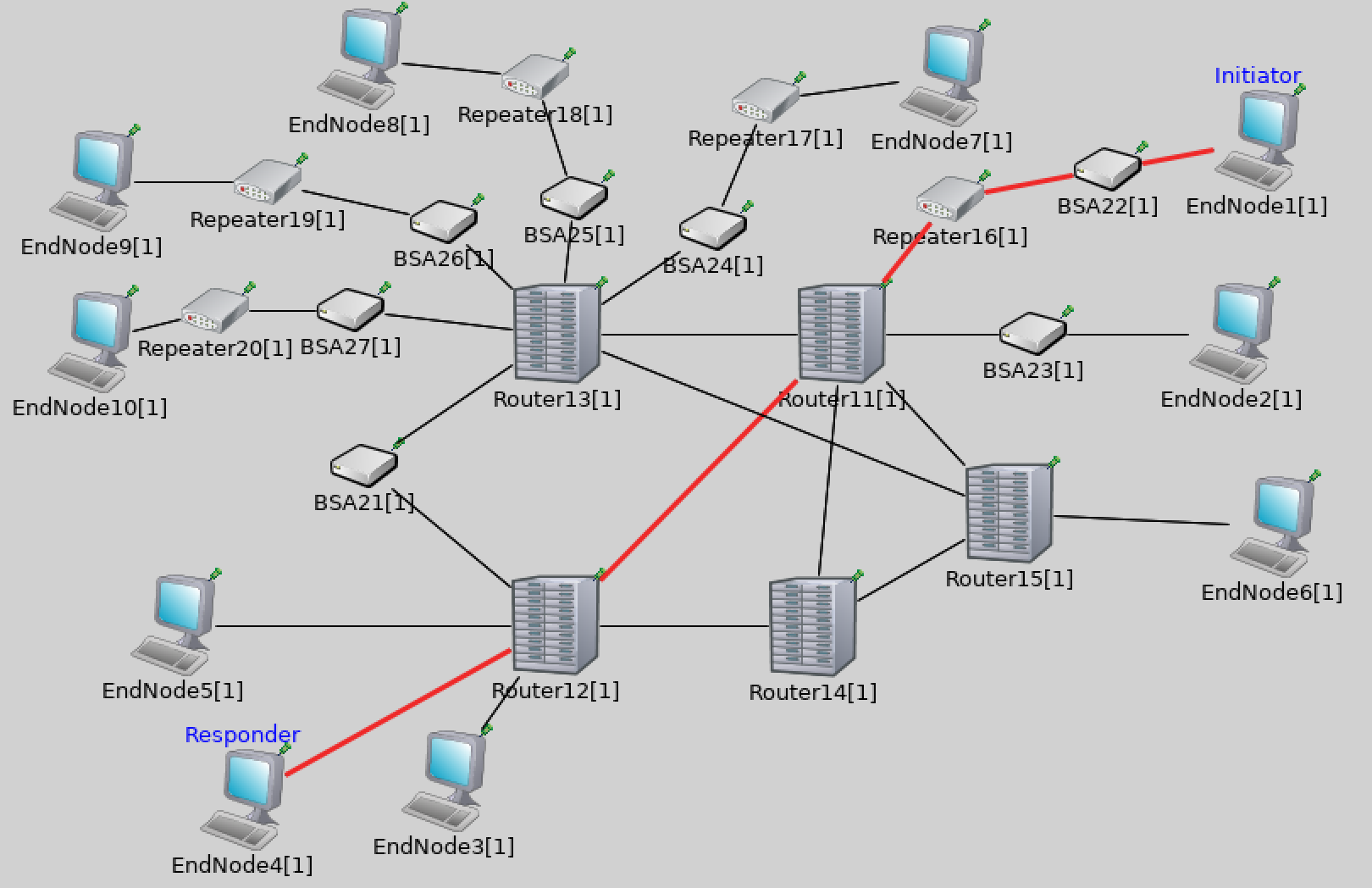}
    \caption{Example network configuration in QuISP, with multiple link types and one connection marked in red. The outermost computer icons represent end nodes that perform applications. Between two end nodes, there are repeaters, routers and BSAs when the link type is MIM.}
    \label{fig: ex_network}
\end{figure}

We demonstrate the workflow to simulate end-to-end entanglement generation between Initiator and Responder on the network shown in Figure~\ref{fig: ex_network}.

First, we prepare a NED file corresponding to the target network. (See Appendix~\ref{sec: ned_definition} for more details.) The nodes
are defined as submodules of the network. After enumerating all involved nodes, we define both classical and quantum connections between these nodes.

\begin{lstlisting}[language=Ini, caption=Example configurations in INI file, label={lst: exnet_ini}]
[Config Example_end_to_end_tomography]
network = Example_Network
sim-time-limit = 15s
seed-set = 1
**.buffers = 100
**.TrafficPattern = 2
**.EndToEndConnection = true
**.distant_measure_count = 8000
**.emission_success_probability = 0.8
**.Measurement_error_rate = 0.0450074
**.Measurement_X_error_ratio = 0.018
**.Measurement_Y_error_ratio = 0.06
**.Measurement_Z_error_ratio = 0.06
\end{lstlisting}

Second, we prepare an INI file. The INI file shown in Listing~\ref{lst: exnet_ini} represents a simulation that is performed on the \verb|Example_Network| topology (\verb|network|) with 100 qubits for each QNIC (\verb|buffers|), and all qubits have some amount of measurement errors (\verb|Measurement_error_rate|).  The success probability of photon emission from one quantum memory is 0.8 (\verb|emission_success_probability|). The paths are generated randomly (\verb|TrafficPattern|), and end nodes measure 8000 Bell pairs for tomography at the end of Rule execution (\verb|distant_measure_count|).
This simulation stops after some amount of time in the simulation (\verb|sim-time-limit|).
Other than these variables, the default values (for example no errors on the link) are used for the simulation.

Once a user compiles the source, the simulation starts. All nodes are booted with the above parameters. QRSA is started, and if indicated tomography on the links (used to populate link information for routing) begins.
For application traffic, each end node acts as an Initiator, randomly selects one other end node to be the Responder, and issues a "Connection Setup Request". For example, for the path highlighted in red in Figure~\ref{fig: ex_network}, the Initiator issues a Connection Setup Request and sends it to the Responder.




The simulation automatically stops and invokes post-processing functions to make statistics. An example output file is shown in Listing~\ref{lst: result}.
Note that the fidelity in the listing is obtained from a density matrix constructed via quantum state tomography run by the application, and therefore corresponds to what real-world software would actually determine from its tests.

\begin{lstlisting}[language=Txt, caption=Result of example simulation between two EndNodes. , label={lst: result}]
EndNode[0]<-->QuantumChannel{
    cost=0.00160205;
    distance=28km;
    fidelity=0.983016;
    bellpair_per_sec=645.955;
    tomography_time=10.83666473934;
    tomography_measurements=7000;
    actual_meas=7000; 
    GOD_clean_pair_total=6887; 
    GOD_X_pair_total=1; 
    GOD_Y_pair_total=1; 
    GOD_Z_pair_total=111;
}<-->EndNode[0]; 
F=0.983016; 
// Error probabilities
X=-0.000188593; 
Z=0.0169845;    
Y=0.000188593;   
...
\end{lstlisting}

%% file: chapters/5.Performance.tex
\section{Correctness and performance}
\label{sec:performance}
Based on the previous discussions on the simulator's design and usage, we now investigate its validity and performance.  First, we give a simple numerical experiment to check whether the result out of the simulator agrees with theoretical values. After that, we extend the scale of the experiment to measure the performance.
\subsection{Correctness}
\label{sec: correctness}

In order to test the correctness of QuISP we set up a small linear network composed of two end nodes A and C, with a single repeater B as depicted in Figure~\ref{fig: entanglement_swapping}.
The end nodes are separated from the repeater by equal distance $d$.

To make this test analytically tractable, we consider a simplified error model for the quantum link by assuming only Pauli $X$ errors may affect the flying qubits with probability $P_X$ per kilometer of the fiber.
This error model may transform the ideal Bell pair $|\Phi^+\rangle = (|00\rangle + |11\rangle) / \sqrt{2}$ into a new state $|\Psi^+\rangle = (|01\rangle + |10\rangle) / \sqrt{2}$ and vice versa.
The state of the two qubits after the flying qubit reaches repeater B is given by a statistical mixture of $|\Phi^+\rangle$ with probability $p_{clean}$, and $|\Psi^+\rangle$ with probability $1 - p_{clean}$,
where
\begin{equation}
    p_{clean} = \frac{1}{2} \left[ 1 + (1 - 2 P_X)^d \right].
\end{equation}
Upon arrival of the flying qubits to the repeater node B they are measured together in the Bell basis.
This results in the desired state $|\Phi^+\rangle$ between qubits A-C if both pairs A-B$_{\text{1}}$ and B$_{\text{2}}$-C are clean or if both suffered from an error \cite{vanmeter2014quantum}.
Therefore the fidelity of qubit pair A-C is given by
\begin{equation}
    F_{AC} = p_{clean}^2 + (1 - p_{clean})^2.
\end{equation}

\begin{figure}
    \centering
    \includegraphics[width=\linewidth]{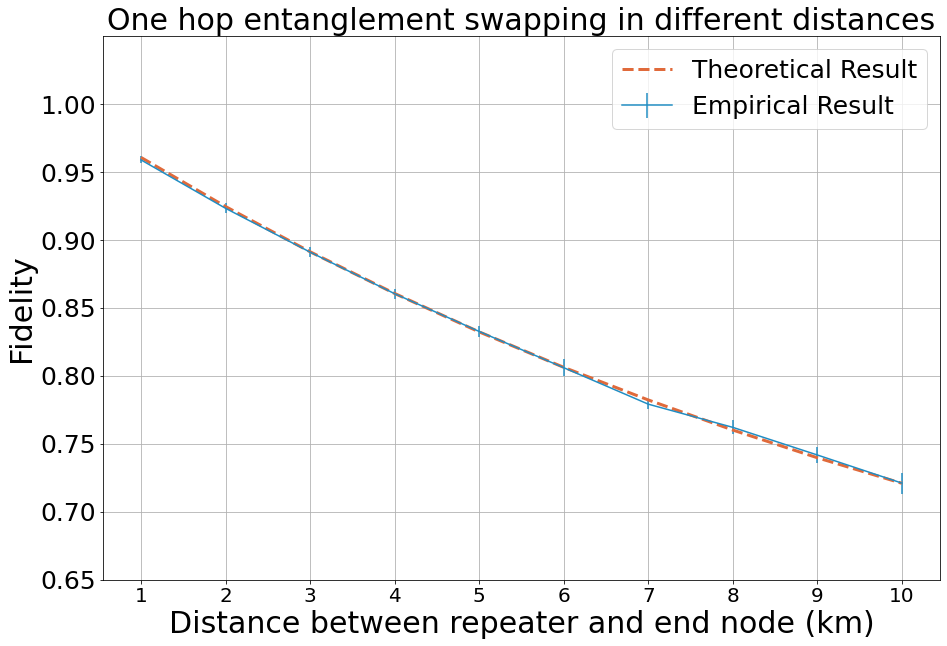}
    \caption{The comparison between theoretical and empirical result of fidelity of the final state for a small network. In the empirical result, the average and standard deviations of 10 data points are depicted as error bars.}
    \label{fig: validation}
\end{figure}

We simulated this network with $P_x = 0.02$ and 7000 measurements. We repeated the simulation ten times. The resulting mean empirical fidelity and its standard deviation are shown in Figure~\ref{fig: validation} alongside the theoretical fidelity $F_{AC}$. We varied the distance $d$ from 1 to 10 kilometers. The empirical result  matches the theoretical prediction. 

\subsection{Scalability}
\label{sec: sim_settings}



In order to show that QuISP can scale up to the goals mentioned in Section~\ref{sec: goals}, we investigated the performance of QuISP in terms of events processed per  second and the duration of CPU time taken to generate one end-to-end Bell pair. 
These results can indicate the relation between the scale of experiments and the time it takes to run the simulation. 
Although the total number of events depends on multiple factors such as the complexity of the protocol, network size, and number of buffer qubits each node has, it can give us some indications regarding the total work required.

\begin{table}[!t]
\centering
\caption{Simulation Environment}
\label{tab: system_spec}
\begin{tabular}{|l|l|}
\hline
Machine  & MacBook Pro 16-inch 2019 model                                                                     \\ \hline
Guest OS & Ubuntu 18.04 on Docker                                                                             \\ \hline
Host OS  & macOS Big Sur (11.5.2)                                                                               \\ \hline
CPU      & \begin{tabular}[c]{@{}l@{}}2.3GHz 8-Core 9th Gen. Intel Core i9\\ (8 cores assigned to Docker)\end{tabular} \\ \hline
Memory   & \begin{tabular}[c]{@{}l@{}}32 GB 2667MHz DDR4\\ (10 GB assigned to Docker)\end{tabular}            \\ \hline
\end{tabular}
\end{table}

For these tests, we decided to use the Docker environment QuISP provides.
Table~\ref{tab: system_spec} details the simulation environment. 

The work required to run the simulation can be broken down into three main parts: routine background task for link generation, connection setup, and RuleSet execution. The total simulation duration will be the aggregation of time required for each connection to establish, and each connection is scaled by number of hops, number of Bell pairs required, and the success probability of link generation.  Since the load of RuleSet execution is directly related to the number of connections for any network size and connection setup is a very lightweight task, we decided to show the performance of QuISP using the simplest case, a linear topology with one connection.

\begin{figure}[!t]
    \centering
    \includegraphics[width=\linewidth]{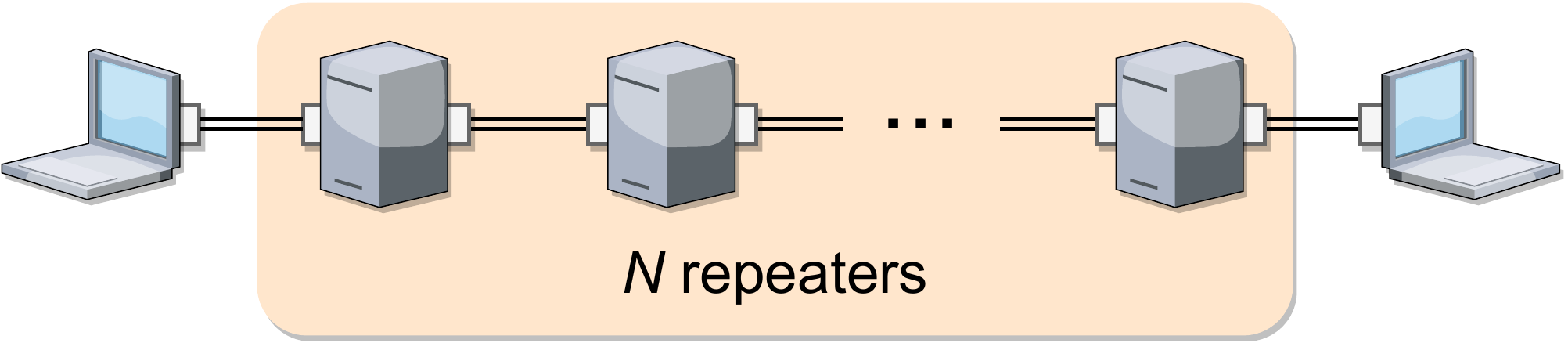}
    \caption{Linear network with different number of repeaters. One repeater has two QNICs and both end nodes have one QNIC.} 
    \label{fig: linear_network}
\end{figure}

Figure~\ref{fig: linear_network} is the network that we use for performance analysis. We performed quantum state tomography using 200 E2E Bell pairs between these two end nodes with the probability of link generation at $0.32$.
We can adjust the number of repeaters between two end nodes, with the E2E distance growing as we add links. Each repeater has two QNICs with $n_Q$ qubits per QNIC. The end nodes contain one QNIC with $n_Q$ qubits.
The total number of qubits is therefore $N = 2 n_Q (n_R + 1)$, where $n_R$ is the number of repeaters.

The RuleSet structure we used in this benchmark is similar to the one depicted in Figure~\ref{fig: ex_ruleset}. The Bell pairs are purified once before undergoing entanglement swapping at every step. The order of the entanglement swapping was chosen to be the binary tree structure~\cite{briegel1998quantum}. The purification is also done on the end-to-end Bell pair shared between the two end nodes before tomography.

\begin{figure}[!t]
    \centering
    \includegraphics[width=\linewidth]{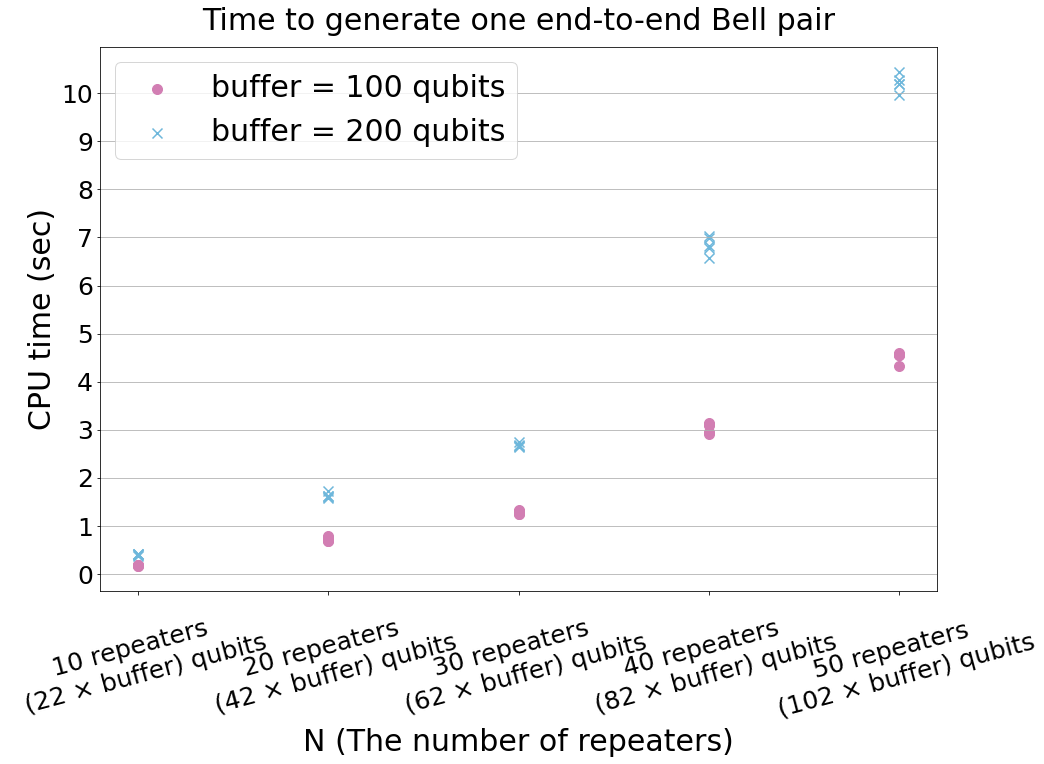}
    \caption{CPU time to generate one end-to-end Bell pair with different number of repeaters. Five data points for each network and the number of buffer qubits with different seed values.
    }
    \label{fig: cputime_bellpair}
\end{figure}

Figure~\ref{fig: cputime_bellpair} shows the average CPU time (in seconds) per end-to-end Bell pair generated. 
When the number of repeaters increases, the amount of time to generate one end-to-end Bell pair increases because it includes far more operations (more entanglement swapping operations and more rounds of purification) than that with the smaller number of repeaters. In theory, the scaling of work required for generating end-to-end Bell pair with $N$ repeaters is linearithmic ($O(N\log N$)). We can clearly see that the scaling of our simulation time grows no worse than polynomially. 

\begin{figure}[]
    \centering
    \includegraphics[width=\linewidth]{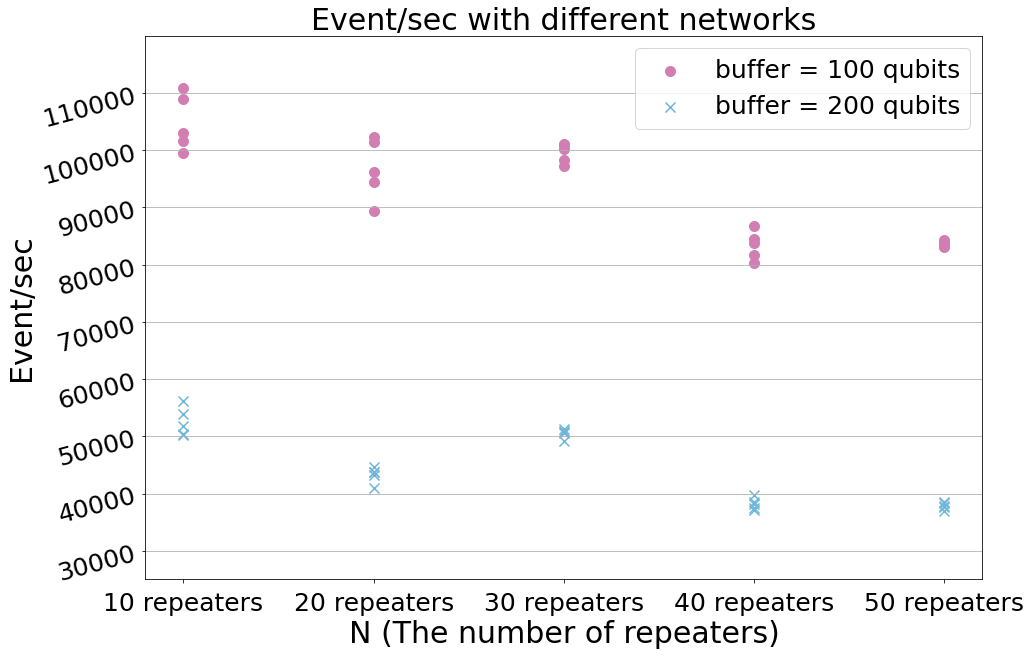}
    \caption{The number of events per second on different networks and with differing numbers of qubits.}
    \label{fig: event_per_sec}
\end{figure}

We also investigated the number of events processed per second. Figure~\ref{fig: event_per_sec} shows the number of events per second for different network sizes and number of qubits inside QNICs. 
Increasing the number of repeaters results in longer simulation time in the scaling as we expected. On the other hand, doubling the buffer qubits inside a QNIC (from 100 to 200) resulted in around twice the simulation time even though the total number of processed events are similar. This indicates that QuISP might have some kind of unintended overhead which scales linearly on the number of buffer qubits, which we expect to fix in a near-term release.


%% file: chapters/6.Conclusion.tex
\section{Conclusion}
\label{sec:conclusion}
This paper proposed a discrete event simulator for a large-scale quantum network on top of OMNeT++. QuISP is based on two essential aspects, the RuleSet protocol and error basis simulation. The RuleSet protocol allows us to establish end-to-end connections that operate fully asynchronously and autonomously. Error basis is a new representation of the quantum state, which only tracks the transition of errors rather than the full quantum state, enabling simulation that scales efficiently in both qubits involved in a single quantum state and in number of network nodes, links and connections.

We provided evidence that suggests QuISP properly models the real world by comparing simulation results with analytic results for a small  network.  We also assessed the performance of QuISP in terms of the actual CPU time and the number of events handled per second. While this cost varies with configuration, simulation cost does not row explosively, and simulations finish fast enough for researchers to explore aspects of the design space smoothly.

The simulator is a work in progress, but is complete enough for use as a research tool, and indeed our group has several projects running in parallel, each of which begins by defining the Rules that govern a new type of network functionality. Our current advanced projects are extending the functionality to include 2G, all-optical, multi-party protocols and internetworking.

Planned improvements can be split into three categories: core functionality, performance, and extensions. Connection teardown is incomplete, and routing, while correctly handling heterogeneity, is static; these limit our ability to compare multiplexing schemes effectively, and will be remedied soon. Performance measurements suggest substantial room to improve, and we expect to reach our goal of simulating an internet of 100 networks of 100 nodes each. 

All of this is, of course, in pursuit of a functional, robust, extensible, and above all useful Quantum Internet. To get there, we must design and test architecture and protocols with emphasis on scale and heterogeneity. We expect that QuISP will contribute to advancing this design and implementation.

%% file: chapters/7.Acknowledgement.tex
\section{Acknowledgement}
This material is based upon work supported by the Air Force Office of Scientific Research under award number FA2386-19-1-4038.

\section*{Availability}
The simulator source code, documentation and sample configurations are fully available online as an open-source project. Feedback, requests for features, and code contributions are welcome. Our mission is not only to build an advanced simulator, but also to make it easy to use for quantum networking research and as a teaching and learning tool. We are also alpha testing a web browser-based simulation interface created by compiling OMNeT++ into Web Assembly (wasm), allowing new users to learn about quantum networking without dealing with the complex build environment of OMNeT++.

%% file: chapters/A.Appendix.tex
\clearpage
\section{NED (Network Definition) in OMNeT++}
\label{sec: ned_definition}

\begin{lstlisting}[language=Ned, caption=Ned file for example network, label={lst:
exned}]
network Example_Network {
    submodules:
        EndNode[11]: QNode {
            address = index;
            nodeType = "EndNode";
        }
        Router[5]: QNode {
            address = 100 + index;
            nodeType = "Router";
        }
        Repeater[6]: QNode {
            address = 200 + index;
            nodeType = "Repeater";
        }
        Hom[8]: HoM {
            address = 300 + index;
        }
    connections:
        Router[0].port++ <--> ClassicalChannel{ distance=19km; } <--> Router[1].port++;
        ...
        Router[0].quantum_port++ <--> QuantumChannel {  distance = 19km; } <--> Router[1].quantum_port_receiver++;
        ...
}
\end{lstlisting}
Network definition (NED) provided by OMNeT++. \verb|network| symbol defines the network with \verb|submodules| listed in it.

Listing~\ref{lst: exned} is an example of NED syntax that defines the network with several quantum devices. Here, we have 11 \verb|EndNode|, 5 \verb|Router|, 6 \verb|Repeater| and 8 \verb|HoM| initialized with vectors of each module. The connections between these components are defined under \verb|connections| which specify the port to be used and the type of channel such as \verb|ClassicalChannel| to connect them.

The devices involved in the network are also defined as NED style language by nesting them as submodules and passing the parameters they require.In this examaple, \verb|address|,  \verb|nodeType| and \verb|distance| are given by network. The atomic components such as BSA, StationaryQubit are defined as \verb|simple|, and the combination of simple modules such as \verb|QNode|, \verb|Router| are defined as \verb|module|.

\section{Transition matrix $Q$}
\label{sec:transition_matrix}
The transition matrix $Q$ allows us to evolve the error probability vector in time and is given by
\begin{widetext}
\begin{equation}
Q =
\begin{pmatrix}
P_I & P_X & P_Y & P_Z & P_R & P_E & P_L \\
P_X & P_I & P_Z & P_Y & P_R & P_E & P_L  \\
P_Y & P_Z & P_I & P_X & P_R & P_E & P_L  \\
P_Z & P_Y & P_X & P_I & P_R & P_E & P_L  \\
0 & 0 & 0 & 0 & P_I+P_Z+P_R & P_X+P_Y+P_E & P_L \\
0 & 0 & 0 & 0 & P_X+P_Y+P_R & P_I+P_Z+P_E & P_L \\
0 & 0 & 0 & 0 & 0 & 0 & 1
\end{pmatrix}, \label{eq:Q_matrix}
\end{equation}
\end{widetext}
where $P_j$ represents the error probability per unit time.
It captures the nature in which quantum errors interact with each other.
For example, a qubit affected by a Pauli $X$ error can be transformed into a clean qubit by another Pauli $X$ error acting on it. This is reminiscent of how two classical bit flip errors cancel each other.
Other errors may interact in more complicated ways. For example, a qubit that has been affected by the energy relaxation error $R$ will stay in the same state if no further error occurs, or a Pauli $Z$ occurs, or a the relaxation process takes place again. This is captured by the transition matrix element $Q_{55} = P_I + P_Z + P_R$.
Some error states cannot be reached from certain error states.
In particular, a qubit that has decohered via the relaxation or excitation process cannot go back to being affected by a simple Pauli error.
This is captured by $Q_{ij} = 0$.
In this sense, the most destructive type of error is photon loss $L$ captured by the fact that $Q_{7j} = 0$ except for $Q_{77} = 1$.
Once a qubit is lost it we cannot recover it.




%% file: main.bbl
\begin{thebibliography}{10}

\bibitem{altepeter2005photonic}
Joseph~B Altepeter, Evan~R Jeffrey, and Paul~G Kwiat.
\newblock Photonic state tomography.
\newblock {\em Advances in Atomic, Molecular, and Optical Physics},
  52:105--159, 2005.

\bibitem{aparicio2011master}
Luciano Aparicio.
\newblock Design and evaluation of communication protocols for quantum repeater
  networks.
\newblock Master's thesis, University of Tokyo, 2011.

\bibitem{arute2019quantum}
Frank Arute, Kunal Arya, Ryan Babbush, Dave Bacon, Joseph~C Bardin, Rami
  Barends, Rupak Biswas, Sergio Boixo, Fernando~GSL Brandao, David~A Buell,
  et~al.
\newblock Quantum supremacy using a programmable superconducting processor.
\newblock {\em Nature}, 574(7779):505--510, 2019.

\bibitem{PRXQuantum.2.017002}
David Awschalom, Karl~K. Berggren, Hannes Bernien, Sunil Bhave, Lincoln~D.
  Carr, Paul Davids, Sophia~E. Economou, Dirk Englund, Andrei Faraon, Martin
  Fejer, Saikat Guha, Martin~V. Gustafsson, Evelyn Hu, Liang Jiang, Jungsang
  Kim, Boris Korzh, Prem Kumar, Paul~G. Kwiat, Marko Lon\ifmmode~\check{c}\else
  \v{c}\fi{}ar, Mikhail~D. Lukin, David~A.B. Miller, Christopher Monroe,
  Sae~Woo Nam, Prineha Narang, Jason~S. Orcutt, Michael~G. Raymer, Amir~H.
  Safavi-Naeini, Maria Spiropulu, Kartik Srinivasan, Shuo Sun, Jelena
  Vu\ifmmode \check{c}\else \v{c}\fi{}kovi\ifmmode~\acute{c}\else \'{c}\fi{},
  Edo Waks, Ronald Walsworth, Andrew~M. Weiner, and Zheshen Zhang.
\newblock Development of quantum interconnects (quics) for next-generation
  information technologies.
\newblock {\em PRX Quantum}, 2:017002, Feb 2021.

\bibitem{bennett1996concentrating}
Charles~H. Bennett, Herbert~J. Bernstein, Sandu Popescu, and Benjamin
  Schumacher.
\newblock Concentrating partial entanglement by local operations.
\newblock {\em Phys. Rev. A}, 53:2046--2052, Apr 1996.

\bibitem{bennett2014quantum}
Charles~H. Bennett and Gilles Brassard.
\newblock Quantum cryptography: Public key distribution and coin tossing.
\newblock {\em Theoretical Computer Science}, 560:7--11, 2014.
\newblock Theoretical Aspects of Quantum Cryptography – celebrating 30 years
  of BB84.

\bibitem{bennett1993teleportation}
Charles~H. Bennett, Gilles Brassard, Claude Cr\'epeau, Richard Jozsa, Asher
  Peres, and William~K. Wootters.
\newblock Teleporting an unknown quantum state via dual classical and
  einstein-podolsky-rosen channels.
\newblock {\em Phys. Rev. Lett.}, 70:1895--1899, Mar 1993.

\bibitem{briegel1998quantum}
H.-J. Briegel, W.~D\"ur, J.~I. Cirac, and P.~Zoller.
\newblock Quantum repeaters: The role of imperfect local operations in quantum
  communication.
\newblock {\em Phys. Rev. Lett.}, 81:5932--5935, Dec 1998.

\bibitem{broadbent2009universal}
Anne Broadbent, Joseph Fitzsimons, and Elham Kashefi.
\newblock Universal blind quantum computation.
\newblock In {\em 2009 50th Annual IEEE Symposium on Foundations of Computer
  Science}, pages 517--526. IEEE, 2009.

\bibitem{chen:1805.01450}
Jianxin Chen, Fang Zhang, Cupjin Huang, Michael Newman, and Yaoyun Shi.
\newblock Classical simulation of intermediate-size quantum circuits, 2018.
\newblock arXiv:1805.01450.

\bibitem{chen2021integrated}
Yu-Ao Chen, Qiang Zhang, Teng-Yun Chen, Wen-Qi Cai, Sheng-Kai Liao, Jun Zhang,
  Kai Chen, Juan Yin, Ji-Gang Ren, Zhu Chen, et~al.
\newblock An integrated space-to-ground quantum communication network over
  4,600 kilometres.
\newblock {\em Nature}, 589(7841):214--219, 2021.

\bibitem{chia2020phase}
Andy Chia, Michal Hajdu\ifmmode~\check{s}\else \v{s}\fi{}ek, Ranjith Nair,
  Rosario Fazio, Leong-Chuan Kwek, and Vlatko Vedral.
\newblock Phase-preserving linear amplifiers not simulable by the parametric
  amplifier.
\newblock {\em Phys. Rev. Lett.}, 125:163603, 2020.

\bibitem{chia2019phase}
Andy Chia, Michal Hajdu{\v{s}}ek, Rosario Fazio, Leong-Chuan Kwek, and Vlatko
  Vedral.
\newblock Phase diffusion and the small-noise approximation in linear
  amplifiers: {L}imitations and beyond.
\newblock {\em {Quantum}}, 3:200, 2019.

\bibitem{coopmans2020netsquid}
Tim Coopmans, Robert Knegjens, Axel Dahlberg, David Maier, Loek Nijsten, Julio
  Oliveira, Martijn Papendrecht, Julian Rabbie, Filip Rozp{\k{e}}dek, Matthew
  Skrzypczyk, Leon Wubben, Walter de~Jong, Damian Podareanu, Ariana
  Torres~Knoop, David Elkouss, and Stephanie Wehner.
\newblock Netsquid, a discrete-event simulation platform for quantum networks.
\newblock {\em arXiv preprint arXiv:2010.12535}, 2020.

\bibitem{cuomo2020towards}
Daniele Cuomo, Marcello Caleffi, and Angela~Sara Cacciapuoti.
\newblock Towards a distributed quantum computing ecosystem.
\newblock {\em IET Quantum Communication}, 1(1):3--8, 2020.

\bibitem{dahlberg2019link}
Axel Dahlberg, Matthew Skrzypczyk, Tim Coopmans, Leon Wubben, Filip
  Rozp{\k{e}}dek, Matteo Pompili, Arian Stolk, Przemys{\l}aw Pawe{\l}czak,
  Robert Knegjens, Julio de~Oliveira~Filho, et~al.
\newblock A link layer protocol for quantum networks.
\newblock In {\em Proceedings of the ACM Special Interest Group on Data
  Communication}, pages 159--173. 2019.

\bibitem{devitt2013quantum}
Simon~J Devitt, William~J Munro, and Kae Nemoto.
\newblock Quantum error correction for beginners.
\newblock {\em Reports on Progress in Physics}, 76(7):076001, 2013.

\bibitem{diadamo2020qunetsim}
Stephen DiAdamo, Janis N{\"o}zel, Benjamin Zanger, and Mehmet~Mert Be{\c{s}}e.
\newblock Qunetsim: A software framework for quantum networks.
\newblock {\em arXiv preprint arXiv:2003.06397}, 2020.

\bibitem{dowling2003quantum}
Jonathan~P Dowling and Gerard~J Milburn.
\newblock Quantum technology: the second quantum revolution.
\newblock {\em Philosophical Transactions of the Royal Society of London.
  Series A: Mathematical, Physical and Engineering Sciences},
  361(1809):1655--1674, 2003.

\bibitem{duan2001long}
L-M Duan, Mikhail~D Lukin, J~Ignacio Cirac, and Peter Zoller.
\newblock Long-distance quantum communication with atomic ensembles and linear
  optics.
\newblock {\em Nature}, 414(6862):413--418, 2001.

\bibitem{duan2010colloquium}
L.-M. Duan and C.~Monroe.
\newblock Colloquium: Quantum networks with trapped ions.
\newblock {\em Rev. Mod. Phys.}, 82:1209--1224, Apr 2010.

\bibitem{dynes2019cambridge}
JF~Dynes, Adrian Wonfor, WW-S Tam, AW~Sharpe, R~Takahashi, M~Lucamarini,
  A~Plews, ZL~Yuan, AR~Dixon, J~Cho, et~al.
\newblock Cambridge quantum network.
\newblock {\em npj Quantum Information}, 5(1):1--8, 2019.

\bibitem{ekert1991quantum}
Artur~K. Ekert.
\newblock Quantum cryptography based on {B}ell's theorem.
\newblock {\em Phys. Rev. Lett.}, 67:661--663, Aug 1991.

\bibitem{elliott2003quantum}
Chip Elliott, David Pearson, and Gregory Troxel.
\newblock Quantum cryptography in practice.
\newblock SIGCOMM '03, page 227–238, New York, NY, USA, 2003. Association for
  Computing Machinery.

\bibitem{fujii2009entanglement}
Keisuke Fujii and Katsuji Yamamoto.
\newblock Entanglement purification with double selection.
\newblock {\em Phys. Rev. A}, 80:042308, Oct 2009.

\bibitem{gottesman97:_thesis}
Daniel Gottesman.
\newblock {\em Stabilizer Codes and Quantum Error Correction}.
\newblock PhD thesis, California Institute of Technology, May 1997.

\bibitem{horodecki2007quantum}
Ryszard Horodecki, Pawe\l{} Horodecki, Micha\l{} Horodecki, and Karol
  Horodecki.
\newblock Quantum entanglement.
\newblock {\em Rev. Mod. Phys.}, 81:865--942, Jun 2009.

\bibitem{zukowski1993event}
M.~\ifmmode~\dot{Z}\else \.{Z}\fi{}ukowski, A.~Zeilinger, M.~A. Horne, and
  A.~K. Ekert.
\newblock ``{E}vent-ready-detectors'' bell experiment via entanglement
  swapping.
\newblock {\em Phys. Rev. Lett.}, 71:4287--4290, Dec 1993.

\bibitem{ilo2018remote}
Ebubechukwu~O Ilo-Okeke, Louis Tessler, Jonathan~P Dowling, and Tim Byrnes.
\newblock Remote quantum clock synchronization without synchronized clocks.
\newblock {\em npj Quantum Information}, 4(1):1--5, 2018.

\bibitem{jiang2007oaq}
L.~Jiang, J.M. Taylor, N.~Khaneja, and M.D. Lukin.
\newblock {Optimal approach to quantum communication using dynamic
  programming}.
\newblock {\em Proceedings of the National Academy of Sciences}, 104(44):17291,
  2007.

\bibitem{PhysRevA.79.032325}
Liang Jiang, J.~M. Taylor, Kae Nemoto, W.~J. Munro, Rodney Van{ }Meter, and
  M.~D. Lukin.
\newblock Quantum repeater with encoding.
\newblock {\em Phys. Rev. A}, 79(3):032325, Mar 2009.

\bibitem{jing2019entanglement}
Bo~Jing, Xu-Jie Wang, Yong Yu, Peng-Fei Sun, Yan Jiang, Sheng-Jun Yang, Wen-Hao
  Jiang, Xi-Yu Luo, Jun Zhang, Xiao Jiang, et~al.
\newblock Entanglement of three quantum memories via interference of three
  single photons.
\newblock {\em Nature Photonics}, 13(3):210--213, 2019.

\bibitem{jones2016design}
Cody Jones, Danny Kim, Matthew~T Rakher, Paul~G Kwiat, and Thaddeus~D Ladd.
\newblock Design and analysis of communication protocols for quantum repeater
  networks.
\newblock {\em New Journal of Physics}, 18(8):083015, 2016.

\bibitem{joshi2020trusted}
Siddarth~Koduru Joshi, Djeylan Aktas, S{\"o}ren Wengerowsky, Martin Lon{\v
  c}ari{\'c}, Sebastian~Philipp Neumann, Bo~Liu, Thomas Scheidl,
  Guillermo~Curr{\'a}s Lorenzo, {\v Z}eljko Samec, Laurent Kling, Alex Qiu,
  Mohsen Razavi, Mario Stip{\v c}evi{\'c}, John~G. Rarity, and Rupert Ursin.
\newblock A trusted node{\textendash}free eight-user metropolitan quantum
  communication network.
\newblock {\em Science Advances}, 6(36), 2020.

\bibitem{jozsa2000quantum}
Richard Jozsa, Daniel~S. Abrams, Jonathan~P. Dowling, and Colin~P. Williams.
\newblock Quantum clock synchronization based on shared prior entanglement.
\newblock {\em Phys. Rev. Lett.}, 85:2010--2013, Aug 2000.

\bibitem{kimble2008quantum}
H~Jeff Kimble.
\newblock The quantum internet.
\newblock {\em Nature}, 453(7198):1023--1030, 2008.

\bibitem{kozlowski00:qnp}
Wojciech Kozlowski, Axel Dahlberg, and Stephanie Wehner.
\newblock Designing a quantum network protocol.
\newblock In {\em Proceedings of the 16th International Conference on Emerging
  Networking EXperiments and Technologies}, CoNEXT '20, page 1–16, New York,
  NY, USA, 2020. Association for Computing Machinery.

\bibitem{I-D.irtf-qirg-principles}
Wojciech Kozlowski, Stephanie Wehner, Rodney~Van Meter, Bruno Rijsman,
  Angela~Sara Cacciapuoti, Marcello Caleffi, and Shota Nagayama.
\newblock Architectural principles for a quantum internet.
\newblock Internet-Draft draft-irtf-qirg-principles-06, IETF Secretariat,
  February 2021.
\newblock
  \url{https://www.ietf.org/archive/id/draft-irtf-qirg-principles-06.txt}.

\bibitem{ladd06:_hybrid_cqed}
T.~D. Ladd, P.~van Loock, K.~Nemoto, W.~J. Munro, and Y.~Yamamoto.
\newblock Hybrid quantum repeater based on dispersive {CQED} interaction
  between matter qubits and bright coherent light.
\newblock 8:184, 2006.

\bibitem{markov:contracting-siam}
Igor~L. Markov and Yaoyun Shi.
\newblock Simulating quantum computation by contracting tensor networks.
\newblock {\em SIAM Journal on Computing}, 38(3):963--981, 2008.

\bibitem{matsuo2019quantum}
Takaaki Matsuo, Cl{\'e}ment Durand, and Rodney Van~Meter.
\newblock Quantum link bootstrapping using a ruleset-based communication
  protocol.
\newblock {\em Physical Review A}, 100(5):052320, 2019.

\bibitem{michie1968memo}
Donald Michie.
\newblock “memo” functions and machine learning.
\newblock {\em Nature}, 218(5136):19--22, 1968.

\bibitem{mirhosseini2020superconducting}
Mohammad Mirhosseini, Alp Sipahigil, Mahmoud Kalaee, and Oskar Painter.
\newblock Superconducting qubit to optical photon transduction.
\newblock {\em Nature}, 588(7839):599--603, 2020.

\bibitem{muralidharan2016optimal}
Sreraman Muralidharan, Linshu Li, Jungsang Kim, Norbert L{\"u}tkenhaus,
  Mikhail~D Lukin, and Liang Jiang.
\newblock Optimal architectures for long distance quantum communication.
\newblock {\em Scientific reports}, 6(1):1--10, 2016.

\bibitem{nagayama21:e2eep}
Shota Nagayama.
\newblock Towards end-to-end error management for a quantum internet.
\newblock Released simultaneously to arXiv., December 2021.

\bibitem{nielsen2002quantum}
Michael~A Nielsen and Isaac Chuang.
\newblock {\em Quantum computation and quantum information}.
\newblock Cambridge University Press, 2002.

\bibitem{omnetpp_hp}
OMNeT++.
\newblock {OMN}e{T}++.
\newblock {\em https://omnetpp.org/}, 2021.

\bibitem{orus2014practical}
Rom{\'a}n Or{\'u}s.
\newblock A practical introduction to tensor networks: Matrix product states
  and projected entangled pair states.
\newblock {\em Annals of Physics}, 349:117--158, 2014.

\bibitem{pathumsoot2020modeling}
Poramet Pathumsoot, Takaaki Matsuo, Takahiko Satoh, Michal
  Hajdu\ifmmode~\check{s}\else \v{s}\fi{}ek, Sujin Suwanna, and Rodney
  Van~Meter.
\newblock Modeling of measurement-based quantum network coding on a
  superconducting quantum processor.
\newblock {\em Phys. Rev. A}, 101:052301, 2020.

\bibitem{pednault:1710.05867}
Edwin Pednault, John~A. Gunnels, Giacomo Nannicini, Lior Horesh, Thomas
  Magerlein, Edgar Solomonik, and Robert Wisnieff.
\newblock {B}reaking the 49-{Q}ubit {B}arrier in the {S}imulation of {Q}uantum
  {C}ircuits, 2017.
\newblock arXiv:1710.05867.

\bibitem{peev2009theSECOQC}
M~Peev, C~Pacher, R~All{\'{e}}aume, C~Barreiro, J~Bouda, W~Boxleitner,
  T~Debuisschert, E~Diamanti, M~Dianati, J~F Dynes, S~Fasel, S~Fossier,
  M~Fürst, J-D Gautier, O~Gay, N~Gisin, P~Grangier, A~Happe, Y~Hasani,
  M~Hentschel, H~Hübel, G~Humer, T~Länger, M~Legr{\'{e}}, R~Lieger,
  J~Lodewyck, T~Lorünser, N~Lütkenhaus, A~Marhold, T~Matyus, O~Maurhart,
  L~Monat, S~Nauerth, J-B Page, A~Poppe, E~Querasser, G~Ribordy, S~Robyr,
  L~Salvail, A~W Sharpe, A~J Shields, D~Stucki, M~Suda, C~Tamas, T~Themel, R~T
  Thew, Y~Thoma, A~Treiber, P~Trinkler, R~Tualle-Brouri, F~Vannel, N~Walenta,
  H~Weier, H~Weinfurter, I~Wimberger, Z~L Yuan, H~Zbinden, and A~Zeilinger.
\newblock The {SECOQC} quantum key distribution network in vienna.
\newblock {\em New Journal of Physics}, 11(7):075001, jul 2009.

\bibitem{pompili2021realization}
M.~Pompili, S.~L.~N. Hermans, S.~Baier, H.~K.~C. Beukers, P.~C. Humphreys,
  R.~N. Schouten, R.~F.~L. Vermeulen, M.~J. Tiggelman, L.~dos Santos~Martins,
  B.~Dirkse, S.~Wehner, and R.~Hanson.
\newblock Realization of a multinode quantum network of remote solid-state
  qubits.
\newblock {\em Science}, 372(6539):259--264, 2021.

\bibitem{proctor2018multiparameter}
Timothy~J. Proctor, Paul~A. Knott, and Jacob~A. Dunningham.
\newblock Multiparameter estimation in networked quantum sensors.
\newblock {\em Phys. Rev. Lett.}, 120:080501, Feb 2018.

\bibitem{omnetpp_repo}
GitHub repository.
\newblock {OMN}e{T}++.
\newblock {\em https://github.com/omnetpp/omnetpp}, 2021.

\bibitem{sasaki2011field}
M.~Sasaki, M.~Fujiwara, H.~Ishizuka, W.~Klaus, K.~Wakui, M.~Takeoka, S.~Miki,
  T.~Yamashita, Z.~Wang, A.~Tanaka, K.~Yoshino, Y.~Nambu, S.~Takahashi,
  A.~Tajima, A.~Tomita, T.~Domeki, T.~Hasegawa, Y.~Sakai, H.~Kobayashi,
  T.~Asai, K.~Shimizu, T.~Tokura, T.~Tsurumaru, M.~Matsui, T.~Honjo, K.~Tamaki,
  H.~Takesue, Y.~Tokura, J.~F. Dynes, A.~R. Dixon, A.~W. Sharpe, Z.~L. Yuan,
  A.~J. Shields, S.~Uchikoga, M.~Legr\'{e}, S.~Robyr, P.~Trinkler, L.~Monat,
  J.-B. Page, G.~Ribordy, A.~Poppe, A.~Allacher, O.~Maurhart, T.~L\"{a}nger,
  M.~Peev, and A.~Zeilinger.
\newblock Field test of quantum key distribution in the tokyo qkd network.
\newblock {\em Opt. Express}, 19(11):10387--10409, 2011.

\bibitem{satoh2021attacking}
Takahiko Satoh, Shota Nagayama, Shigeya Suzuki, Takaaki Matsuo, Michal
  Hajdu\v{s}ek, and Rodney~Van Meter.
\newblock Attacking the quantum internet.
\newblock {\em IEEE Transactions on Quantum Engineering}, 2:1--17, 2021.

\bibitem{stucki2011long}
D~Stucki, M~Legr{\'{e}}, F~Buntschu, B~Clausen, N~Felber, N~Gisin, L~Henzen,
  P~Junod, G~Litzistorf, P~Monbaron, L~Monat, J-B Page, D~Perroud, G~Ribordy,
  A~Rochas, S~Robyr, J~Tavares, R~Thew, P~Trinkler, S~Ventura, R~Voirol,
  N~Walenta, and H~Zbinden.
\newblock Long-term performance of the {SwissQuantum} quantum key distribution
  network in a field environment.
\newblock {\em New Journal of Physics}, 13(12):123001, 2011.

\bibitem{vanmeter2014quantum}
Rodney Van~Meter.
\newblock {\em Quantum Networking}.
\newblock John Wiley \& Sons, 2014.

\bibitem{van-meter07:banded-repeater-ton}
Rodney Van{ }Meter, Thaddeus~D. Ladd, W.~J. Munro, and Kae Nemoto.
\newblock System design for a long-line quantum repeater.
\newblock {\em {IEEE/ACM} Transactions on Networking}, 17(3):1002--1013, June
  2009.

\bibitem{I-D.van-meter-qirg-quantum-connection-setup}
Rodney Van{ }Meter and Takaaki Matsuo.
\newblock Connection setup in a quantum network.
\newblock Internet-Draft draft-van-meter-qirg-quantum-connection-setup-01, IETF
  Secretariat, September 2019.

\bibitem{van-meter21:aqia}
Rodney Van{ }Meter, Ryosuke Satoh, Naphan Benchasattabuse, Takaaki Matsuo,
  Michal Hajdu{\v{s}}ek, Takahiko Satoh, Shota Nagayama, and Shigeya Suzuki.
\newblock A quantum internet architecture.
\newblock Released simultaneously to arXiv., December 2021.

\bibitem{varga2001discrete}
Andr{\'a}s Varga.
\newblock Discrete event simulation system.
\newblock In {\em Proc. of the European Simulation Multiconference
  (ESM’2001)}, pages 1--7, 2001.

\bibitem{varga2010omnet++}
Andras Varga.
\newblock Omnet++.
\newblock In {\em Modeling and tools for network simulation}, pages 35--59.
  Springer, 2010.

\bibitem{wehner2018quantum}
Stephanie Wehner, David Elkouss, and Ronald Hanson.
\newblock Quantum internet: A vision for the road ahead.
\newblock {\em Science}, 362(6412), 2018.

\bibitem{wootters1982single}
William~K Wootters and Wojciech~H Zurek.
\newblock A single quantum cannot be cloned.
\newblock {\em Nature}, 299(5886):802--803, 1982.

\bibitem{wu21:sequence}
Xiaoliang Wu, Alexander Kolar, Joaquin Chung, Dong Jin, Tian Zhong, Rajkumar
  Kettimuthu, and Martin Suchara.
\newblock Sequence: a customizable discrete-event simulator of quantum
  networks.
\newblock {\em Quantum Science and Technology}, 2021.

\bibitem{yu2020entanglement}
Yong Yu, Fei Ma, Xi-Yu Luo, Bo~Jing, Peng-Fei Sun, Ren-Zhou Fang, Chao-Wei
  Yang, Hui Liu, Ming-Yang Zheng, Xiu-Ping Xie, et~al.
\newblock Entanglement of two quantum memories via fibres over dozens of
  kilometres.
\newblock {\em Nature}, 578(7794):240--245, 2020.

\bibitem{zhong2020quantum}
Han-Sen Zhong, Hui Wang, Yu-Hao Deng, Ming-Cheng Chen, Li-Chao Peng, Yi-Han
  Luo, Jian Qin, Dian Wu, Xing Ding, Yi~Hu, et~al.
\newblock Quantum computational advantage using photons.
\newblock {\em Science}, 370(6523):1460--1463, 2020.

\bibitem{zhuang2019physical}
Quntao Zhuang and Zheshen Zhang.
\newblock Physical-layer supervised learning assisted by an entangled sensor
  network.
\newblock {\em Phys. Rev. X}, 9:041023, Oct 2019.

\end{thebibliography}
